\documentclass[a4paper,11pt]{article}

\usepackage{amsmath}
\usepackage{amssymb}
\usepackage{amsthm}
\usepackage{graphicx}
\usepackage{caption}
\usepackage{subcaption}
\usepackage{footnote}

\newcommand{\beq}{\begin{equation}}
\newcommand{\eeq}{\end{equation}}

\newcommand{\ba}{\begin{array}}
\newcommand{\ea}{\end{array}}

\newcommand{\bea}{\begin{eqnarray}}
\newcommand{\eea}{\end{eqnarray}}

\newcommand{\bc}{\begin{center}}
\newcommand{\ec}{\end{center}}

\newcommand{\ds}{\displaystyle}

\newcommand{\bt}{\begin{tabular}}
\newcommand{\et}{\end{tabular}}

\newcommand{\bi}{\begin{itemize}}
\newcommand{\ei}{\end{itemize}}

\newcommand{\bd}{\begin{description}}
\newcommand{\ed}{\end{description}}

\newcommand{\p}{\partial}
\newcommand{\sech}{\mbox{sech}}

\newcommand{\cf}{{\it cf.}~}

\newcommand{\comment}[1]{}

\title{\bf Conservation Laws and Web-Solutions for the Benney--Luke Equation}
\author{}
\date{}
\begin{document}
\maketitle

\begin{center}
{\bf Mark J. Ablowitz \\ \bf Christopher W. Curtis \footnote[1]{Corresponding Author: christopher.w.curtis@colorado.edu}} \\ Department of Applied Mathematics \\ University of Colorado at Boulder\\ Boulder, CO, 80309, USA
\end{center}

\begin{abstract}
A long wave multi-dimensional approximation of shallow water waves is the bi-directional Benney--Luke equation. It yields the well-known Kadomtsev--Petviashvili equation in a quasi one-directional limit.   A direct perturbation method is developed; it uses the underlying conservation laws  to determine the slow evolution of parameters of two space dimensional, non-decaying  web-type solutions to the Benney--Luke equation.  New numerical simulations, based on windowing methods which are effective for non-decaying data,  are presented. These simulations  support the analytical results and elucidate the relationship between the Kadomtsev--Petviashvilli and the Benney--Luke equations and are also used to obtain amplitude information regarding particular web solutions.  Additional dissipative perturbations to the Benney--Luke equation are also studied.
\end{abstract}

\section{Introduction}
The modeling of the propagation of waves on the surface of water is a classical problem and remains an active and important area of research.  However, the mathematical formulation of the problem, also known as the Euler water wave equations, presents a number of difficulties such as strong nonlinearity and an unknown location of the boundary that makes finding  solutions challenging.  To deal with this problem, researchers have employed a number of approximate models to the full system of equations describing surface flow.  

In weakly nonlinear water waves whose depth is shallow with respect to the wavelength of the wave, a maximally-balanced multi-dimensional approximation to the Euler water wave equations is the Kadomtsev--Petviashvilli (KP) equation.  There are two signs associated with the KP equation depending on surface tension (ST); these are termed the KP-I (large ST) and KP-II equations (small ST).  Here we assume surface tension is sufficiently small such that the KP-II equation is the relevant approximation to the Euler water wave equations, \cf \cite{aseg}.  Further, in order to derive the KP equation, one also assumes that the height of the wave is small compared to the depth, and that the flow is quasi one-dimensional; {\it i.e.} transverse variations  are slower than those in the primary direction of propagation ({\it cf.}  \cite{abl} for a derivation and details).  While other two-dimensional, shallow-water approximations to the Euler equations can be found, the KP equation is the unique model that is maximally-balanced.  This, and the fact that it  has a Lax Pair and many known closed form solutions makes the KP equation an  important model.

In recent years, a novel class of solutions, called `web-solutions', of the KP equation have been discovered and analyzed.  Building on work in \cite{satsuma} and the resonant solutions found in \cf \cite{milesone}, \cite{milestwo}, in \cite{kod} the web solutions were obtained in terms of Wronskians via Sato's formalism \cite{ohta}. Important properties and pertinent theorems concerning web-solutions can be found in \cite{chak}.  Likewise, experimental and observational evidence exists, \cf  \cite{AblWW}, \cite{kodtwo}, \cite{yeh}, that indicates that these web-solutions are present and persistent in nature.   

From a mathematical perspective it is important to understand whether these solutions are robust to perturbations. In this regard, it is useful to analyze improved approximations to the Euler equations which themselves yield the KP equation.  An approximation that is the object of study in this paper is the Benney--Luke (BL) equation, \cf \cite{benney}.  The BL equation is also a shallow water approximation to the Euler equation which arises naturally as an approximation to water waves.   However, unlike the KP equation, the BL equation allows for two-directional  waves.  Also important in our study is an asymptotically equivalent modification of the BL equation, which unlike the KP equation, has a dispersion relationship in which the group velocity of arbitrarily large wave numbers is bounded; we consider this a BL type equation as well.  The latter BL equation, aside from being a two-directional approximation to the Euler equations,  is expected to be well posed over a broader class of initial conditions than the KP equation.   

Previous studies of the Benney--Luke equation have appeared in the literature.  The analysis in \cite{milesone} begins with the BL equation, and a variant of the Benney--Luke equation was studied numerically in the context of the Mach-reflection problem in \cite{funa}.  In each of these papers, particular web-solutions, so-called `Y-junction' solutions,  were studied.  While this paper addresses the case of small surface tension in the BL equation, we note that the large surface tension case has also been studied.  Rigorous results establishing the existence of lump soliton solutions to the BL equation have appeared, \cf \cite{peg}, and the dynamics of these lump solitons have been investigated, \cf \cite{berger}.  

Recently, we developed a  perturbation theory of web solutions of the BL equation  based techniques from the theory of integrable systems. We established in \cite{ac} that the web-solutions of the KP equation persist for asymptotically long periods of time in the BL equation.  This implies that the web-solutions are robust to the higher-order effects introduced by the BL equation.

The purpose of the present paper is to: i)  develop a direct perturbation method based on conservation laws of the BL equation and  to ii) present new computations of the BL equation which are consistent with web type initial data initial data that is non-decaying.  In one dimension the conservation law approach is widely used \cf \cite{ak}. This is  due to its inherent simplicity and does not require one to employ more complex integrable system methods.  However, in two dimensions since conserved quantities are expressed in terms of integrals, and the web-solutions do not decay in one of the spatial directions the integrals are infinite. To deal with this difficulty, we present a modified definition of the conserved quantities of KP that works for the web-solutions.  

With this modification, the conservation law approach now provides a direct method to perform perturbation analysis (via multiple scales methods) on the web-solutions. As indicated above it does not require integrable systems machinery and readily achieves the same results.  Using the convenience of this conservation law approach, we are also able to analyze the impact of a typical type of dissipation on web-solutions in the Benney--Luke equation.  We chose a linear local dissipative term which creates `shelves' behind the web solution. In one dimension, due to mean term interactions, small amplitude shelves of long extent are known to arise from KdV models with such a dissipative term \cf \cite{ak}.

While the type of dissipation we study is particular, the method we present can, in principle, be used for many other, more physically realistic, dissipation models, \cf \cite{dutykh}.  In this paper our aim is not to decide on the best dissipative model. Instead, we explain how the direct conservation law approach needs to be modified when dissipation is present.  We further exhibit the growth of shelves from the linear local dissipative model.  This method can also be extended to higher order approximations of water waves. However, carrying out this effort is outside the scope of this paper.

The asymptotic methods in this paper rely on separating the web-solutions into what we define as near and far field components.  Using our conservation law based asymptotic method, the far field components are studied analytically.  The near field component is not required in the leading order perturbation analysis.  However, the near field is where the various parts of a web-solution interact.  Therefore, we employ new numerical simulations of the BL equation to investigate the near field, or interaction region.  In this paper, we in particular look at both Y-junctions and another class of web-solutions, the X-waves (or O-type solutions in the terminology of \cite{chak}).  

We also examine the Y-junction solution further by studying the amplification of the so called ``Mach stem", \cf \cite{milestwo}. By varying the small parameter representing the `shallowness' of the water and the angle of opening of the `Y', we show how the relative amplification (see Section \ref{figuressection}) of the Mach stem in the BL equation varies.  We find that the relative amplitude, or amplification ratio, of the stem increases as the water becomes shallower and this result holds uniformly over several different angles.  As the water becomes  shallower, or as the BL equation tends to the KP equation, the amplification ratios of the BL equation tend increase to those of the KP equation.  For the KP Y-junction solution, the largest amplification ratio is known to be four \cite{milestwo}. For the BL equation we find that this  ratio is reduced by $O(\epsilon)$. An interesting open question is: how much is this ratio reduced in the full water wave equations?

Thus the numerical study yields results consistent with asymptotic arguments and is also useful for predictive purposes.  We also numerically solve the BL equation with our chosen dissipation term and show that small shelves of long extent  form in the wake of the Y-junction; this is also consistent with previous one dimensional results, \cf \cite{ak}.  All of these results are new in the literature.

In summary, in this paper we: 
\begin{itemize}
\item Develop a new, conservation law based perturbation method applicable to any web solution.  This significantly simplifies previous, integrable systems based approaches, and thus allows us to model perturbations to the KP equation more readily.  
\item Examine, for the first time, the role dissipation has on web-solutions.  While we have chosen a particular dissipation model, our methods are, in principle, applicable to any dissipation model.  
\item Present new numerical results showing how higher order shallow water effects influence the evolution of the interaction region of web-solutions.  In the case of Y-junctions, we do this over a large number of parameter values and investigate relative amplification ratios; these results are consistent with KP theory.  
\end{itemize}
While in this paper we only study web-solutions in the context of shallow water flow, we believe that the combination of asymptotic and numerical methods presented here will be useful in other nonlinear problems involving non-decaying profiles.   

\subsection{The Benney--Luke Equation and Web-Solutions of the KP Equation}

The Benney-Luke (BL) equation is given in non-dimensional form by \cite{abl} 
\begin{equation}
q_{tt} - \tilde{\Delta}q + \alpha \epsilon \tilde{\Delta}^{2}q + \epsilon(\p_{t}|\tilde{\nabla}q|^{2}+q_{t}\tilde{\Delta}q)=0,
\label{bl}
\end{equation}
where small dispersion, slow transverse variation and weak nonlinearity are all balanced.  These small effects are denoted by $\epsilon, |\epsilon| \ll 1$; also $q$ is the velocity potential, $\tilde{\nabla}=\left<\p_{x},\epsilon^{1/2}\p_{y}\right>$, $\tilde{\Delta} = \p^{2}_{x} + \epsilon\p^{2}_{y}$, and $\alpha = \tilde{\sigma}-\frac{1}{3}$, where $\tilde{\sigma}$ is related to the surface tension.  Since the BL equation is a long wave approximation, as an initial value problem, it can and does  suffer from arbitrarily large growth rates. 

By replacing $\partial_t^2$ by $\partial_x^2$ we shall use a a regularized version of the BL equation which is asymptotically equivalent to the above BL equation (\ref{bl}) and is written in the form
\begin{equation}
q_{tt} - \tilde{\Delta}q + \alpha \epsilon \tilde{\Delta}q_{tt} + \epsilon(\p_{t}|\tilde{\nabla}q|^{2} + q_{t}\tilde{\Delta}q) = 0.
\label{blreg}
\end{equation}
We use the transformation $\xi = x - t$, $\tau = \epsilon t$, so that 
\[
\p_{\xi} = \p_{x}, ~ \p_{t} = -\p_{\xi} + \epsilon \p_{\tau}. 
\]
In this paper we only consider small surface tension, hence  $\tilde{\sigma} \sim 0$, so that $\alpha \sim -1/3$ and thus this regularized BL equation is linearly well-posed; it has the dispersion relationship $\omega^{2} = \frac{|\tilde{k}|^{2}}{1+\epsilon|\alpha||\tilde{k}|^{2}}$, where $\tilde{k} = \left<k_{x},\sqrt{\epsilon}k_{y} \right>$.

Introducing the scaling $A = (-8\alpha)^{1/5}, ~ \beta = \frac{2}{A^{2}}, ~ \gamma^{2} = \frac{6}{A^{3}}, ~ \delta = - \frac{2}{A},$ and letting $q = \frac{A}{\beta}\phi(\beta \xi, \gamma y, \delta \tau)$,  \eqref{blreg} becomes
\[
\ba{l}
4\phi_{\xi \tau} - 3 \phi_{yy} - \phi_{\xi\xi\xi\xi} - 6\phi_{\xi}\phi_{\xi\xi} + \epsilon A \left(2\phi_{\tau\tau} - 2 \phi_{\xi\xi\xi\tau} - \frac{3}{2}\phi_{\xi\xi y y} - 6\phi_{y}\phi_{\xi y} \right. \\
\\
\left. - 4\phi_{\xi}\phi_{\xi\tau} - 3\phi_{\xi}\phi_{yy} - 2\phi_{\tau}\phi_{\xi \xi} \right) = 0.
\ea 
\]
Defining the operator (see \cite{ablclar})
\[
\p^{-1}_{\xi}u(\xi,y,t) = \frac{1}{2}\int_{-\infty}^{\xi} u(z,y,t) dz - \frac{1}{2}\int_{\xi}^{\infty} u(z,y,t) dz,
\]
and  letting $u = \phi_{\xi}$ and $\tilde{\epsilon}=\epsilon A$ the BL equation we consider is given by 
\begin{equation}
(-4u_{\tau} + u_{\xi \xi \xi}  + 6uu_{\xi})_{\xi} + 3u_{yy} - \tilde{\epsilon} \p_{\xi}F(u) = 0, 
\label{rescalbl}
\end{equation}
where $F(u)$ is given by
\[
\ba{rl}
F(u)   = & 2\p^{-1}_{\xi}u_{\tau \tau} - 2u_{\xi \xi \tau} - \frac{3}{2}u_{\xi y y} - 3 \p_{\xi}(\p^{-1}_{\xi}u_{y})^{2} \\
&\\
&- 4uu_{\tau} - 3u\p^{-1}_{\xi}u_{yy} - 2u_{\xi}\p^{-1}_{\xi}u_{\tau}.
\ea
\]
Hereafter we replace $\tilde{\epsilon}$ by $\epsilon$  (also note that for small surface tension $A\sim1.2$).  The definition of $\p^{-1}_{\xi}$ is chosen so that if $f, g \in L^{1}(\mathbb{R})$, then $\int_{\mathbb{R}}f(\xi)\p^{-1}_{\xi}g(\xi)d\xi$ is well defined, and 
\[
\int_{\mathbb{R}}f(\xi)\p^{-1}_{\xi}g(\xi)d\xi = - \int_{\mathbb{R}}g(\xi)\p^{-1}_{\xi}f(\xi)d\xi.
\]
Alternatively the perturbed BL equation is written as 
\begin{equation}
u_{\tau}= K(u) -  \frac{\epsilon}{4} F(u)
\label{pertbl}
\end{equation}
where $K(u)=\frac{1}{4}u_{\xi\xi\xi} + \frac{3}{2} uu_{\xi} + \frac{3}{4} \p^{-1}_{\xi} u_{yy}$ .

As seen from the rescaled BL equation (\ref{rescalbl}), the KP equation gives the leading order behavior of the BL equation.  We now introduce the web-solutions to KP which serve as the particular leading order behavior of interest.  A web solution, say $w(\xi,y,\tau)$ to the KP equation in Wronskian form is given by (cf. \cite{chak}; note for the general N-soliton solution in Hirota form see \cite{satsuma})
\beq
w(\xi,y,\tau) = 2 \p^{2}_{\xi}\log(\Omega(\xi,y,\tau)), 
\label{omega}
\eeq
where $\Omega(\xi,y,\tau)=\mbox{Wr}(f_{1},\cdots,f_{N})$, where $\mbox{Wr}$ denote the Wronskian of the functions $f_{i}$ (\cf \cite{chak}).  The functions $f_{i}$ have the particular form
\beq
f_{i} = \sum_{j=1}^{M} b_{ij} e^{\theta_{j}},
\label{f_in_Wr}
\eeq
with $b_{ij}$ constant, $N < M$, and $\theta_{j}(\xi,y,\tau) = k_{j}\xi +k^{2}_{j}y+k^{3}_{j}\tau$, where $k_{1}< \cdots < k_{n}$.  It was proved in \cite{chak} that as $y\rightarrow \infty$, there are $N$ lines, and as $y \rightarrow -\infty$,  $M-N$ lines, in the $\xi,y$ plane with slopes $c_{ij}=-(k_{i}+k_{j})^{-1}$, $j>i$, such that 
\beq
w \sim \frac{1}{2}(k_{j}-k_{i})^2\sech^{2}\left(\frac{\theta_{j}-\theta_{i}+\theta_{ij}}{2} \right). \label{asymkp}
\eeq
where the phases $\theta_{ij}$ are constant in this pure KP solution. 

We now introduce a perturbation scheme for studying the behavior of solutions $u(\xi,y,\tau)$ to the BL equation of the form
\[
u(\xi,y,\tau) = w(\xi,y,\tau,T) + \epsilon s(\xi,y,\tau) + \cdots,
\]
where $T=\epsilon \tau$.  The leading order term $w$ is a web-solution to the KP equation in which we allow the terms $k_{j}$ to vary in the slow time $T$.  Motivated by the form of the web solution, it is convenient to separate  the solution $w$  into far field ($|y|\gg 1$) and near field ($|y|=\mathcal{O}(1)$) components (see also \cite{ac}).  The far-field coordinates along a given ray are denoted as $X_{ij}$ and $Y_{ij}$, where $j>i$, and $X_{ij}$ is determined by the differential equations  
\beq
\p_{\xi}X_{ij} = 1, ~ \p_{y} X_{ij} = k_{i}+k_{j}, ~ \p_{\tau} X_{ij} = k^{2}_{j}+ k^{2}_{i} + k_{i} k_{j}, 
\label{xcoord}
\eeq
and $Y_{ij}$ is given by 
\beq
Y_{ij} = y - (k_{j}+k_{i})\xi, 
\label{ycoord}
\eeq
where in the large $|y|$ limit derivatives with respect to $Y_{ij}$ are considered asymptotically negligible.  We also require that $k_{i}(T)+k_{j}(T) = C_{ij}$, where $C_{ij}$ is independent of $\tau$ and $T$; hence $\p_{T} X_{ij}=0$.  This requirement means there is no secular growth in $\partial_Tw$ for large $y$.  

We assume along $|Y_{ij}|\rightarrow \infty$ that 
\[
u \sim w(X_{ij},T) + \epsilon s(X_{ij},\tau)+\cdots.
\]  
In this limit, the equation for $w$ is, 
\[
\p_{X_{ij}}\left(-4\frac{(k^{3}_{j}-k^{3}_{i})}{k_{ij}} w_{X_{ij}} + \p^{3}_{X_{ij}} w + 6ww_{X_{ij}}\right) + 3 (k_{j}+k_{i})^{2}\p^{2}_{X_{ij}}w = 0
\]
where $k_{ij}=k_{j}-k_{i}$.  Integrating once in $X_{ij}$, we get  
\[
-k^{2}_{ij}w_{X_{ij}}+  \p^{3}_{X_{ij}}w + 6ww_{X_{ij}} = 0
\]
which has the soliton solution
\beq
w(X_{ij},T) = \frac{(k_{ij})^{2}}{2}\sech^{2}\left(\frac{k_{ij}(X_{ij} + \theta^{(0)}_{ij}(T))}{2}\right), 
\label{asympsoln}
\eeq
and which is consistent with the previous asymptotic arguments.  The functions $k_{ij}=k_{ij}(T)$ are functions of the slow time and $\theta^{(0)}_{ij}(T)$ is an arbitrary phase; these functions are determined later in the perturbation scheme. 

At the next order,  noting for convenience that we drop the subscripts on $X_{ij}$, the equation for $s(X,\tau)$ is
\[
-4s_{\tau} - k^{2}_{ij}s_{X} + \p^{3}_{X} s + 6(ws)_{X} = \p_{X}\tilde{F}(X) + 4w_{T},
\]
It is also noted that 
\begin{align}
\tilde{F}(X) = & 2\frac{(k^{3}_{j}-k^{3}_{i})^{2}}{k^{2}_{ij}}w(X) - \left(\frac{9}{2}(k_{i}+k_{j})^{2}+3\frac{(k^{3}_{j}-k^{3}_{i})}{k_{ij}} \right)\p^{2}_{X}w \label{redforce}\\
&  -\left(2\frac{(k^{3}_{j}-k^{3}_{i})}{k_{ij}}+\frac{3}{2}(k_{j}+k_{i})^{2}\right)w^{2}(X), \nonumber
\end{align}
and
\beq
w_{T} = 2\frac{\p_{T}k_{ij}}{k_{ij}} w + \left(\frac{\p_{T}k_{ij}}{k_{ij}}(X+\theta^{(0)}_{ij})+\p_{T}\theta^{(0)}_{ij} \right)w_{X}.
\label{slowpertevol}
\eeq
Since $\p_{X}\tilde{F}(X) + 4w_{T}$ is independent of the fast time $\tau$, we separate the perturbation $s$ into $s=h(X)+g(X,\tau)$, 
so that 
\begin{align}
- k^{2}_{ij}h_{X} + \p^{3}_{X} h + 6(wh)_{X} = & \p_{X}\tilde{F}(X) + 4w_{T}, \label{pertforce}\\
-4g_{\tau} - k^{2}_{ij}g_{X} + \p^{3}_{X} g + 6(wg)_{X} = & 0, \nonumber
\end{align}
with $g(X,0)=-h(X)$.  

The slowly varying phases $\theta^{(0)}_{ij}(T)$ were determined in \cite{ac} via the additional orthogonality condition 
\beq
\int_{\mathbb{R}}w(X) h(X) dX = 0. 
\label{seccond}
\eeq
which removed unbounded growth as $\tau \rightarrow \infty$.  We note that orthogonality conditions such as these usually correspond to physical constraints, {\it i.e.} conservation laws. 

 In Section 2,  from the conservation law approach a direct and physically motivated derivation of the condition (\ref{seccond}) is given; also in Section 2 the role of dissipation is examined.  In Section 3, we present our numerical findings concerning the evolution of several different types of web-solutions.  These results indicate that web-solutions are robust to these perturbative effects over asymptotically significant lengths of time.  The numerical study of the amplification of the stem of a Y-Junction in the BL equation, and the effect of dissipation, is also presented in Section 3.  The Appendix collects some of the technical details of the approach used in Sections 2 and 3.     

\section{Conservation Law Scheme for Arbitrary Web-Solutions}

In order to determine the evolution of the slow coefficients, we use a method (see  \cite{ak}), which employs the conserved quantities of the leading order integrable problem to determine the evolution of slow variables.  We use conservation of energy which we define by the following iterated integral
\[
\mathcal{E}(u) = \frac{1}{2}  \left<  \int_{\mathbb{R}} u^{2} d\xi \right>_{y}  = \lim_{L\rightarrow \infty} \frac{1}{2L}\int^{L}_{-L}\int_{\mathbb{R}}  u^{2}(\xi,y,\tau) d\xi dy . 
\]
The average along the $y$-axis is taken because the line solutions to the KP equation do not decay along the $y$-axis.  It is also important to point out that for web-solutions and their perturbations that 
\[
\lim_{L\rightarrow \infty} \int_{\mathbb{R}} \frac{1}{L}\int^{L}_{-L} u^{2}(\xi,y,\tau) dy d\xi \neq \int_{\mathbb{R}} \lim_{L\rightarrow \infty}  \frac{1}{L}\int^{L}_{-L} u^{2}(\xi,y,\tau) dy d\xi,
\] 
and so the placement of the limit is important.  

Using the above definition of energy it follows that 
\[
\p_{\tau} \mathcal{E}(u)  =\left<  \ds{ \int_{\mathbb{R}}} u u_{\tau} \ d\xi \right>_{y} =  \left<  \ds{ \int_{\mathbb{R}}} u K(u) d\xi \right>_{y} - \frac{\epsilon}{4} \left< \ds{\int_{\mathbb{R}}}u F(u)d\xi \right>_{y}.
\]
The analytic issues concerned with differentiating the energy functional are discussed in the Appendix of this paper.  From the definition of $K(u)$, we expect  that 
\beq
\left< \ds{\int_{\mathbb{R}}} uK(u) d\xi \right>_{y}=0, 
\label{avgcond}
\eeq
which would then give
\beq
\left< \ds{\int_{\mathbb{R}}} u u_{\tau}d\xi \right>_{y}=  -\frac{\epsilon}{4}  \left<  \ds{\int_{\mathbb{R}}} u F(u)d\xi \right>_{y}.
 \label{pseceq}
\eeq
Note, due to the average, in order to show (\ref{avgcond}), we must establish that 
\beq
 \left<  \int_{\mathbb{R}} u\p^{-1}_{\xi}u_{yy} d\xi \right>_{y} = 0. 
\label{finalconclusion}
\eeq
To show this, some care must be used to accommodate for the presence of the average which makes integration by parts more complicated.  This is worked out in the Appendix for web solutions to the KP equation.  We further assume it holds for the perturbation $u = w + \epsilon s + \cdots$, so that taking (\ref{avgcond}) to be true is valid to the order of the asymptotics.  

Since we are assuming $u(\xi,y,\tau)$ is, in particular regions of the plane, a solution of the form
\beq
u(\xi,y,\tau) = w(X_{ij},T) + \epsilon s(X_{ij},\tau) + \mathcal{O}(\epsilon^{2}), 
 \label{exp}
\eeq
and is otherwise at most $\mathcal{O}(\epsilon)$ outside the regions of interest, the ansatz for u implies $w_{\tau} =\epsilon w_T$. 
Further, based on the one dimensional problem it is natural to assume that 
\beq
\left< \ds{\int_{\mathbb{R}}} ws_{\tau}d\xi \right>_{y}= o(\epsilon),
\label{seceq1}
\eeq
(this is discussed in more detail below), then we get  
\beq
\left< \ds{\int_{\mathbb{R}}} w (w_T + \frac{1}{4}F(w)) d\xi \right>_{y} = 0.
\label{seceq2}
\eeq
   
\begin{figure}[!h]
\centering
\includegraphics[width=.30\textwidth]{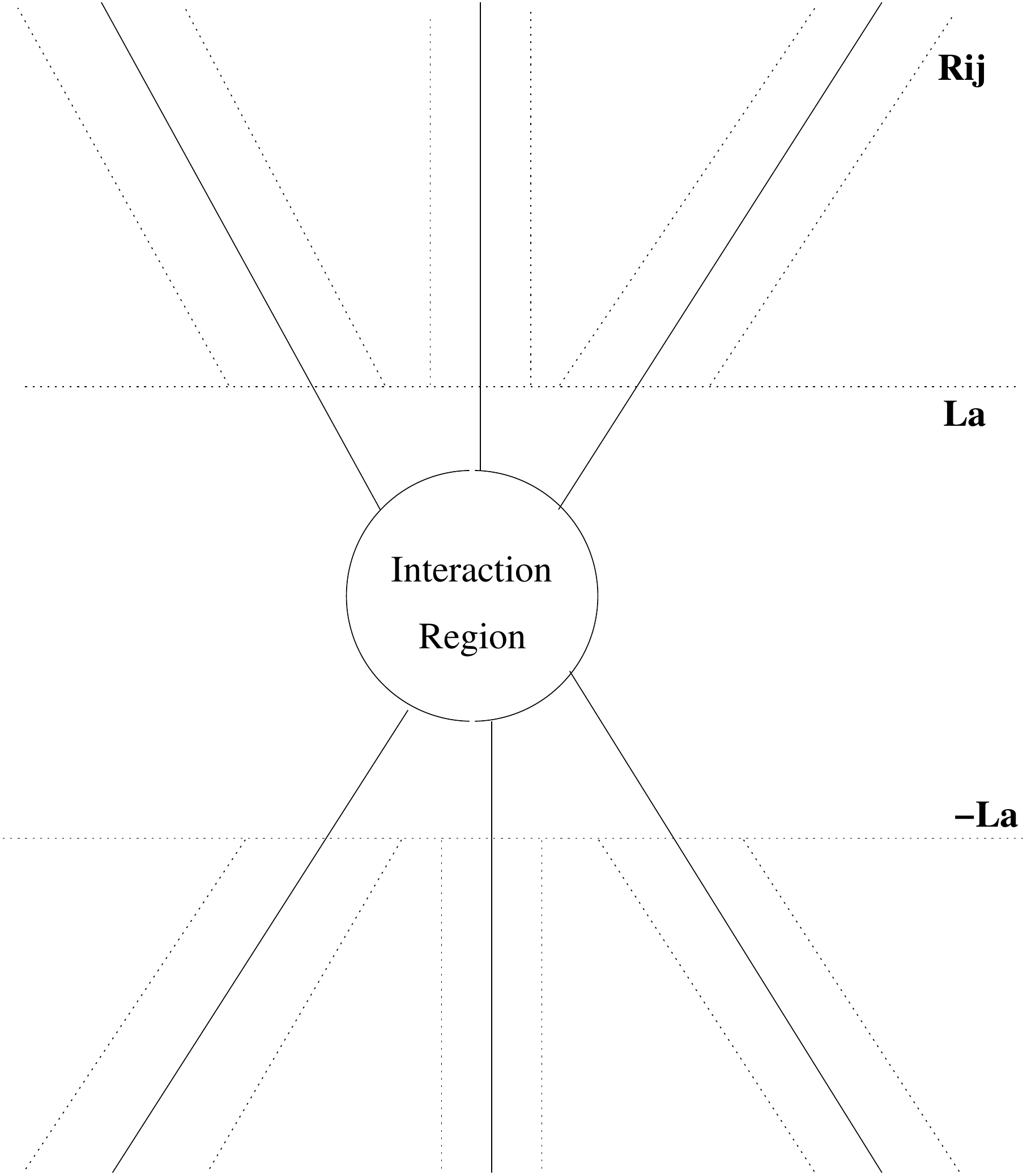}
\caption{Far-Field Region}
\label{fig:farfield}
\end{figure}

We now demonstrate, in some detail, how to use the solvability conditions associated with \eqref{pertforce}) in order to reduce \eqref{pseceq} to \eqref{seceq1} and \eqref{seceq2}.  As $|y| \rightarrow \infty$,  using the coordinate change given by (\ref{xcoord}) and (\ref{ycoord}), $w$ depends only on $X_{lj}$, hence becoming stationary with respect to $\tau$.  The perturbation $s(X_{ij},\tau)$ is written as $s(X_{ij},\tau) = h(X_{ij}) + g(X_{ij},\tau)$ in region $R_{ij}$ (see Fig (\ref{fig:farfield})) where $h$ solves 
\beq
K'(w)h = \frac{1}{4}F(w) + w_{T}. 
\label{statpart}
\eeq
Note, (\ref{statpart}) is the same as (\ref{pertforce}). 
Then in $R_{ij}$, where we ignore the $Y_{ij}$ coordinate and thus reduce the problem to one dimesnion, {\it i.e.} $X_{ij}$, as shown in the introduction, we have $ \left(K'(w)\right)^{\dag}w = 0$.  In order to ensure that a solution $h$ exists to (\ref{statpart}), the Fredholm alternative enforces to leading order that 
\beq
 \int_{\mathbb{R}} w\left(w_T+\frac{1}{4}F(w)\right) dX_{ij} = 0.
\label{locfredcond}
\eeq
This result gives us immediately that the terms $k_{ij}$ are independent of the slow time $T$.  To show this, we first note that 
\beq
\int_{\mathbb{R}} w(X_{ij}) F(X_{ij}) dX_{ij} = \int_{\mathbb{R}} w(X_{ij})\p_{X_{ij}}\tilde{F}(X_{ij}) dX_{ij} = 0,
\label{usefulcondition}
\eeq
since $w$ is even in $X$,  and from (\ref{redforce}), so is $\tilde{F}$. Using (\ref{locfredcond}) then shows that 
\[
\int_{\mathbb{R}} w w_{T}dX_{ij} = 0.
\]
On the other hand, using (\ref{slowpertevol}), we have
\[
\int_{\mathbb{R}} ww_{T} dX_{ij} = \frac{3}{2}\frac{\p_{T}k_{ij}}{k_{ij}} \int_{\mathbb{R}} w^{2}(X_{ij}) dX_{ij}.
\]
Since $\int_{\mathbb{R}} w^{2}(X_{ij}) dX_{ij}\neq 0$, this implies that $\p_{T}k_{ij}=0$.  Having now shown that the coefficients $k_{ij}$ do not vary slowly, and given our assumption that $k_{j}+k_{i}$ does not depend on $T$,  evidently the coefficients $k_{j}$ are also independent of the slow time $T$. 

We now present an argument that establishes (\ref{seceq2}) using (\ref{locfredcond}).  First, we have the identity
\begin{align*}
\int_{-L}^{L}\int_{\mathbb{R}} w\left(w_{T} +\frac{F(w)}{4}\right)d\xi dy = & \int_{-L_{a}}^{L_{a}}\int_{\mathbb{R}} w\left(w_{T} +\frac{F(w)}{4} \right)d\xi dy  \\
& + \int_{-L}^{-L_{a}}\int_{\mathbb{R}} w\left(w_{T} +\frac{F(w)}{4}\right) d\xi dy  \\
& +  \int_{L_{a}}^{L}\int_{\mathbb{R}}w\left(w_{T} +\frac{F(w)}{4}\right) d\xi dy. 
\end{align*}
Taking $L_{a} \gg 1$, and letting $L\rightarrow \infty$, we can write the integrals 
\[
\ds{\int^{\infty}_{L_{a}}\int_{\mathbb{R}}}w\left(w_{T} +\frac{F(w)}{4}\right) dy d\xi, ~ \ds{\int^{-L_{a}}_{-\infty}\int_{\mathbb{R}}}w\left(w_{T} +\frac{F(w)}{4}\right)  dy d\xi 
\]
as sums over the regions $R_{ij}$ since in between the regions, $w$ and its derivatives are exponentially small.  Then, over any region $R_{ij}$, we have
\[
\int\int_{R_{ij}}w\left(w_{T} +\frac{F(w)}{4}\right)  d\xi dy \sim  J_{ij}\int^{\infty}_{Y_{a}}\int_{\mathbb{R}}w\left(w_{T} +\frac{F(w)}{4}\right) dX_{ij}dY_{ij},
\]  
where $J_{lj}=1+(k_{l}+k_{j})^{-2}$ is the Jacobian of the coordinate transformation and $Y_{a}$ is some constant value of $Y_{lj}$ that only depends on the choice of $L_{a}$. Using (\ref{locfredcond}) then establishes that 
\[
\int\int_{R_{ij}} w\left(w_{T} + \frac{F(w)}{4}\right)d\xi dy \sim 0. 
\]
Given that
\[
\lim_{L\rightarrow \infty} \frac{1}{L} \int_{-L_{a}}^{L_{a}}\int_{\mathbb{R}} w\left(w_{T} +\frac{F(w)}{4}\right)  d\xi dy = 0,
\]
since $L_{a}$ is constant, we finally have
\[
\lim_{L \rightarrow \infty} \frac{1}{L}\int_{-L}^{L}\int_{\mathbb{R}} w\left( w_{T} +\frac{F(w)}{4}\right)d\xi dy \sim 0,
\]
or that (\ref{seceq2}) holds.  

We now turn to determining the phases $\theta^{(0)}_{ij}(T)$.  To do this, using (\ref{seceq2}), (\ref{pseceq}) becomes, noting also that $s(\xi,y,0)=0$, 
\beq
 \left<  \int_{\mathbb{R}} s w d\xi \right>_{y} = 0.
\label{globphasecond}
\eeq
This global relation can be separated into near and far-field components;  note the integrals in the interaction region $[-L_a,L_a]$  are negligible in comparison with the far field. Then in the far-field, we can write this integral over regions $R_{ij}$ (again see Figure \ref{fig:farfield} for clarification).  If we assume that after a short time the dominant part of $s(X_{ij},\tau)$ is given by $h(X_{ij})$, then since $k_{j}$ is constant $w_{T} = \p_{T}\theta^{(0)}_{ij}w_{X_{ij}}$, so that (\ref{pertforce}) becomes
\[
-k^{2}_{ij}h + \p^{2}_{X_{ij}}h + 6wh = \tilde{F}(X_{ij}) + \p_{T}\theta^{(0)}_{ij}w.
\]
One gets from the global condition (\ref{globphasecond}) that in $R_{ij}$, again ignoring $Y_{ij}$,
\beq
\int_{\mathbb{R}} h(X_{ij})w(X_{ij})dX_{ij} = 0.
\label{localphasecond}
\eeq   

This is exactly the orthogonality condition derived in \cite{ac}, {\it i.e.} \eqref{seccond}, that was used to determine the evolution of the difference between phases.  To compute this phase evolution, first introduce the transformation $\tilde{X} =  k_{ij}(X_{ij}+\theta^{(0)}_{ij})/2$, and then let $q(\tilde{X})=2\sech^{2}(\tilde{X})$, which turns (\ref{pertforce}) after one integration in $\tilde{X}$ into
\[
-4h + \p^{2}_{\tilde{X}}h +6qh =c^{1}_{i,j}q + c^{2}_{i,j}q^{2} + c^{3}_{i,j}\p^{2}_{\tilde{X}}q
\]
where the coefficients $c^{p}_{i,j}, p=1,2,3$, which implicitly depend on the region of interest $R_{i,j}$, are given by 
\[
\ba{c}
c^{1}_{i,j} = \ds{2\left(\frac{k^{3}_{j}-k^{3}_{l}}{k_{ij}}\right)^{2}+ 4\p_{T}\theta^{(0)}_{ij} }, ~ c^{2}_{i,j} =  \ds{- \frac{3k^{2}_{ij}}{4}\left(\frac{5}{2}k^{2}_{j}+4k_{i}k_{j}+\frac{5}{2}k^{2}_{i}\right)}, \\
\\
c^{3}_{i,j} = \ds{- \frac{k^{2}_{ij}}{4}\left(\frac{7}{2}k^{2}_{j}+5k_{i}k_{j}+\frac{7}{2}k^{2}_{i}\right)}.
\ea
\]
This leads to the solution
\[
h(\tilde{X}) =-2q(\tilde{X})\left(\frac{c^{1}_{i,j}}{8}(-1+\tilde{X}\tanh(\tilde{X})) -\frac{c^{2}_{i,j}}{6}+\frac{c^{3}_{i,j}}{2}\tilde{X}\tanh(\tilde{X})\right).
\]
Using (\ref{localphasecond}) gives $c^{1}_{i,j}=\frac{8}{3}(-2c^{2}_{i,j}/3+c^{3}_{i,j}/2)$ or 
\beq
\p_{T}\theta^{(0)}_{ij} =\frac{( k_{i}^4 - 28 k_{i}^3 k_{j} - 54 k_{i}^2 k_{j}^2 - 28 k_{i}k_{j}^3 + k_{j}^4)}{24}.
\label{phase}
\eeq
With this, we now can compute the speed, $v_{i,j}$, of each part of the far field via the formula
\begin{equation}
v_{i,j}=\p_{\tau}X_{ij} + \epsilon \p_{T}\theta_{ij}^{0},
\label{speedformula}
\end{equation}
where $\p_{\tau}X_{ij}$ is given in \eqref{xcoord}.

Therefore, using a conservation law approach, we have seen how the results of \cite{ac} can be derived in a way that makes no use of integrable systems.  This makes the perturbation method presented in this paper more broadly applicable than that presented in \cite{ac}.         

\subsection{Dissipation}


The Benney--Luke equation conserves energy, and is a Hamiltonian system (see \cite{peg}).  Of course, real physical systems such as water waves are not perfectly conservative.  So it is useful and instructive to study the effect of adding typical damping terms to the conservative model.  While the dissipative model we use is special, it is chosen to show how to deal with methodological difficulties created by dissipation models. It is also instructive to study this dissipation model since it causes the formation of small amplitude shelves with long extent.

For the Benney-Luke equation, we add a  linear local term which leads us to study  an equation of the form
\[
-4u_{\tau} + u_{\xi\xi\xi} + 6uu_{\xi} + 3 \p^{-1}_{\xi}u_{yy} - \epsilon \gamma u - \epsilon F(u) = 0, 
\]
where the $F_j$ are dispersive terms and the constant   $\gamma > 0$  represents the magnitude of damping.  We can then define the function $\tilde{F}_{1}(u) = \gamma u + F(u)$ and repeat the analysis from above.   Matching terms of $\mathcal{O}(\epsilon)$ after expanding $\p_{\tau}\mathcal{E}(u)$ gives
\[
 \left<  \int_{\mathbb{R}} ww_{T} d\xi \right>_{y} +    \p_{\tau}  \left<  \int_{\mathbb{R}} ws d\xi \right>_{y}  = -\frac{1}{4}  \left<  \int_{\mathbb{R}} (\gamma w^{2} + wF(w))d\xi \right>_{y}, 
\]
and repeating the Fredholm alternative argument that led to (\ref{seceq2}), yields
\beq
\left<  \int_{\mathbb{R}} w w_{T} d\xi \right> _y = -\frac{1}{4}  \left<  \int_{\mathbb{R}} (\gamma w^{2} +  w F(w) )d\xi \right>_y . 
\label{massdisprelation}
\eeq
Now we seperate (\ref{massdisprelation}) over the regions $R_{ij}$.  We showed in the previous section, {\it i.e.} Equation (\ref{usefulcondition}), that 
\[
\int_{\mathbb{R}} w F(w) dX_{ij} = 0.
\]

Then, if we naively use the asymptotic form of the solution given by (\ref{asympsoln}) and the corresponding representation for $w_T$  in $R_{ij}$, we get 
\[
\frac{3}{2}\frac{\p_{T}k_{ij}}{k_{ij}} \int_{\mathbb{R}} w^{2}(X_{ij}) dX_{ij} = -\frac{\gamma}{4}\int_{\mathbb{R}}w^{2}(X_{ij})dX_{ij}, 
\]
so that we have
\[
\frac{d k_{ij}}{dT} = -\frac{\gamma}{6} k_{ij}. 
\]
Coupling this condition with the requirement that $k_{i}(T)+k_{j}(T)=C_{ij}$ creates a difficulty.  For example, if we have $k_{1}$, $k_{2}$, and $k_{3}$, then a web-solution whose slopes are determined by say $k_{1}$ and $k_{2}$, and $k_{1}$ and $k_{3}$, would have two different values for $k_{3}(T)$.  Thus we see that the ansatz given by equation (\ref{asympsoln}) for the leading order behavior is not appropriate globally. 

To deal with these inconsistencies, we use a modified ansatz for $w$ (the index $i,j$ on $w$ and $\psi_{l}$, $l=1,2,3$, is understood) of the form
\[
w = \frac{\eta^{2}(T)}{2}\sech^{2}\left(\frac{\eta(T)\left(\bar{X}+\theta^{(0)}_{ij}\right)}{2}\right)
\]
where $\eta(T) = (k_{j}(0)-k_{i}(0))e^{-\gamma T/6}$, and
\[
\p_{\xi}\bar{X}=1, ~ \p_{y}\bar{X} = k_{j}(0)+k_{i}(0), ~ \p_{\tau} \bar{X} = \psi^{2}_{1} + \psi_{1}\psi_{2}+\psi^{2}_{2},
\]
with
\[
\psi_{1}(T) =  \ds{\frac{1}{2}\left(k_{j}(0)+k_{i}(0) - \eta(T) \right)}, ~ \psi_{2}(T) = \ds{\frac{1}{2}\left(k_{j}(0)+k_{i}(0) + \eta(T) \right)}.
\]
Note, the functions $\eta(T)$, $\psi_{1}(T)$, and $\psi_{2}(T)$ vary in each region $R_{ij}$.    

To determine $\theta^{(0)}_{ij}$, again noting that $s(\xi,y,0)=0$, we employ the global relationship
\[
 \left<  \int_{\mathbb{R}} ws d\xi \right>_{y} = 0.
\]
To make use of this identity, we first assume that on $R_{ij}$ that the dominant contribution to $s$ is given by the solution to the stationary equation
\beq
-\eta^{2}h_{\bar{X}} + \p^{3}_{\bar{X}}h + 6(wh)_{\bar{X}} = \p_{X}\tilde{F}(\bar{X}) + 4w_{T}+\gamma w, 
\label{quasistatglobrelat}
\eeq
which is the same equation as (\ref{pertforce}) except now taking the dissipation into account.  Using the transformations from the previous section we solve for $h$ and derive the phase equation
\[
\p_{T}\theta^{(0)}_{ij} =\frac{\gamma}{6\eta}+\frac{(\psi_{1}^4 - 28 \psi_{1}^3 \psi_{2} - 54 \psi_{1}^2 \psi_{2}^2 - 28 \psi_{1}\psi_{2}^3 + \psi_{2}^4)}{24}.
\]

\section{Computation of the Benney-Luke Equation \label{figuressection}}
We first note that the Benney-Luke equation is second order in time. It is convenient for the numerics to transform to a system by introducing the variable $v = u_{\tau}$.  In order to simulate the Benney-Luke equation numerically, we use a windowing method developed in \cite{schlat}. This method introduces a smooth function of compact support, say $V(y)$, such that $V(y)$ is nearly one over some interval in $y$, say $[-L_{y}+\delta, L_{y} - \delta]$, and $V$ has support in $[-L_{y},L_{y}]$.  We use the function 
\[
V(y) = e^{\log(eps) |\frac{y}{L_{y}}|^{20}}, 
\]
where $eps$ is on the order of $10^{-16}$ (approximately machine precision) and write the solution to the Benney-Luke equation as 
\[
\left( \ba{c} u \\ v \ea \right) = V(y)\left( \ba{c} u \\ v \ea \right)+ (1-V(y))\left( \ba{c} u \\ v \ea \right). 
\] 
We use the leading order asymptotic solution computed in the previous sections to evaluate $u$ and $v$ in the far-field with distances  greater than $\mathcal{O}(|L_y|)$ for times $\mathcal{O}(1/\epsilon)$. We denote these far field solutions as $u_{asy}$ and $v_{asy}$.  

We now wish to determine $V(y) u$.  Define the functions $u_{nr} = Vu$, $v_{nr} = Vv$, so that
\[ 
u\sim u_{nr}+ (1-V)u_{asy}, ~ v\sim v_{nr}+(1-V)v_{asy}.
\]
Using the fact that $u_{asy}$ satisfies the KP equation in the far field, and keeping only terms through $O(\epsilon)$, by substituting $u$ and $v$ into the BL equation, it then follows that $u_{nr}$ satisfies the equation
\[
\p_{\tau} \left( \ba{c} u_{nr} \\ v_{nr} \ea \right) = \mathcal{BL}(u_{nr},v_{nr}) + F_{f}(u_{asy},v_{asy}; u_{nr}), 
\]   
where $\mathcal{BL}(u)$ denotes the Benney-Luke equation.  The term $F_{f}$ is given by 
\[
\frac{1}{2\epsilon}\left(\ba{c} 0 \\  \p^{2}_{\xi}(6(1-V)u_{nr}u_{asy}-3V(1-V)u^{2}_{asy}) -6V'\p_{y}u_{asy}-3V''u_{asy} \ea \right) 
\]
which has support only in the intervals $[-L_{y},-L_{y} + \delta]$ and $[L_{y}-\delta,L_{y}]$.  Note, we ignore the window effects from the nonlinearities of $\mathcal{O}(\epsilon)$ since these terms are smaller and isolated to have support in the region $[-L_{y},-L_{y} + \delta]$ and $[L_{y}-\delta,L_{y}]$.  If we work on a large enough domain, the effect of the error should be nominal far from the boundary since the error propagates with finite speed.   

The functions $u_{nr}$ and $v_{nr}$ satisfy at $\tau = 0$ the boundary conditions 
\[
u_{nr}(\xi,-L_{y},0)=u_{nr}(\xi,L_{y},0)=0.  
\]
At later times, we enforce periodic boundary conditions in the $y$-coordinate and choose the window large enough to ensure that 
\[
u_{nr}(\xi,-L_{y},\tau)=u_{nr}(\xi,L_{y},\tau)\sim 0.  
\]
Given the exponential decay in $\xi$, we also restrict the domain in $\xi$ to the interval $[-L_{\xi},L_{\xi}]$ and enforce periodic boundary conditions in $\xi$ and $y$ in order to numerically solve for $u_{nr}$.  This gives us the advantage of being able to use pseudo-spectral methods.  

However, we have to find the operator $\p^{-1}_{\xi}$ in terms of Fourier series.  Writing a typical periodic function, say $\tilde{u}$, as 
\[
\tilde{u}(\xi,y,\tau) = \sum_{j,l=-\infty}^{\infty} a_{jl}(\tau)e^{ \pi i\left(l \xi/L_{\xi} + j y/L_{y}\right)}, 
\]
we calculate the inverse derivative of $\tilde{u}$ via
\begin{align}
\label{dinverseI}
\p^{-1}_{\xi} \tilde{u} =&  \frac{L_{\xi}}{i\pi} \sum_{j}\sum_{l \neq 0} \frac{a_{jl}(\tau)}{l}  e^{ \pi i\left(l \xi/L_{\xi} + j y/L_{y}\right)} + \frac{L_{\xi}}{i\pi} \sum_{j}\sum_{l \neq 0}(-1)^{l+1} \frac{a_{jl}(\tau)}{l}  e^{ \pi i j y/L_{y}} \nonumber\\
& + ~\xi \sum_{j} a_{j0}(\tau) e^{\pi i j y/L_{y}}, 
\end{align}
where we have used $\p^{-1}_{\xi}=\frac{1}{2}\int_{-L_{\xi}}^{\xi} - \frac{1}{2}\int_{\xi}^{L_{\xi}}$.  In the Appendix, we show that the pseudo-spectral representation of the $(j,l)$ mode of the inverse derivative is given by 
\begin{align}
\label{dinverseII}
(\p^{-1}_{\xi} \tilde{u})^{\hat{}}_{j, l} = & (1-\delta_{l 0})\frac{L_{\xi}}{i\pi l}~ \tilde{a}_{j l}(\tau) - \delta_{l 0}\frac{L_{\xi}}{i\pi N_{T}} \sum_{k=-\tilde{N}+1,k\neq 0}^{\tilde{N}} \frac{\tilde{a}_{j k}(\tau)}{k} \nonumber \\
& +  ~(1-\delta_{l 0 })\frac{2L_{\xi} }{N_{T}}\frac{\tilde{a}_{j 0}}{e^{-2\pi i  l/N_{T}} -1} - \delta_{l 0}\frac{L_{\xi}}{N_{T}}. 
\end{align}
where $\tilde{a}_{jl}(\tau) = (-1)^{l+j}a_{jl}(\tau) $ and $N_{T}$ is the number of modes used in the pseudo-spectral approximation.  

\subsection{Evolution of the Y-Junction and X-Wave in the Benney--Luke Equation}
The following figures show top-down surface plots of the numerical evolution of various web-solution profiles.  The time stepping algorithm used was the ETDRK4 method \cite{tref}, which is a variant of the 4th order Runge-Kutta method.    

\begin{figure}[!h]
\centering
\begin{subfigure}[b]{.49\textwidth}
\includegraphics[width=1\textwidth]{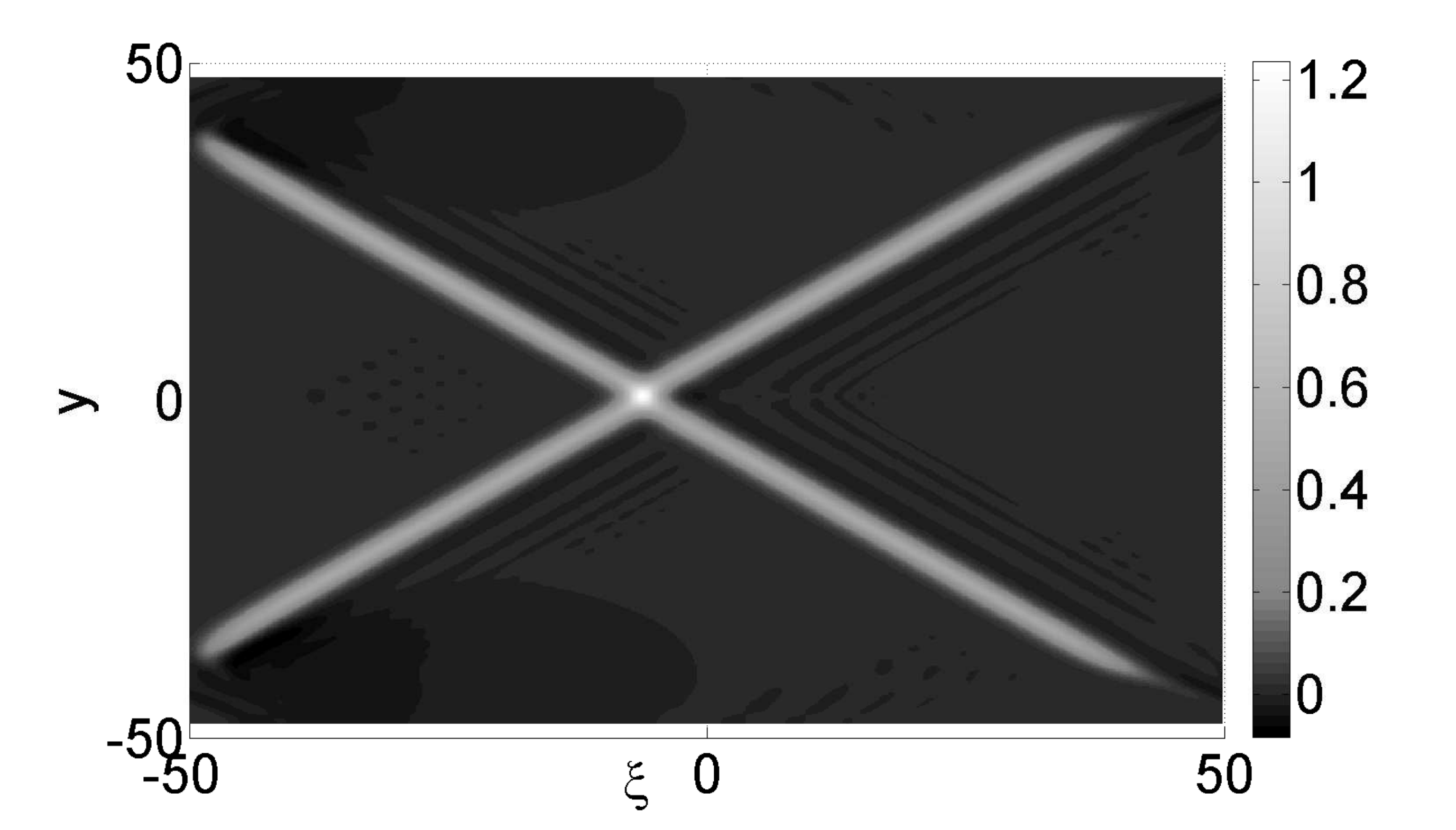}
\caption{Short Stem, $k_{3}=10^{-1}$, $\epsilon = .2$}
\end{subfigure}
\begin{subfigure}[b]{.49\textwidth}
\includegraphics[width=1\textwidth]{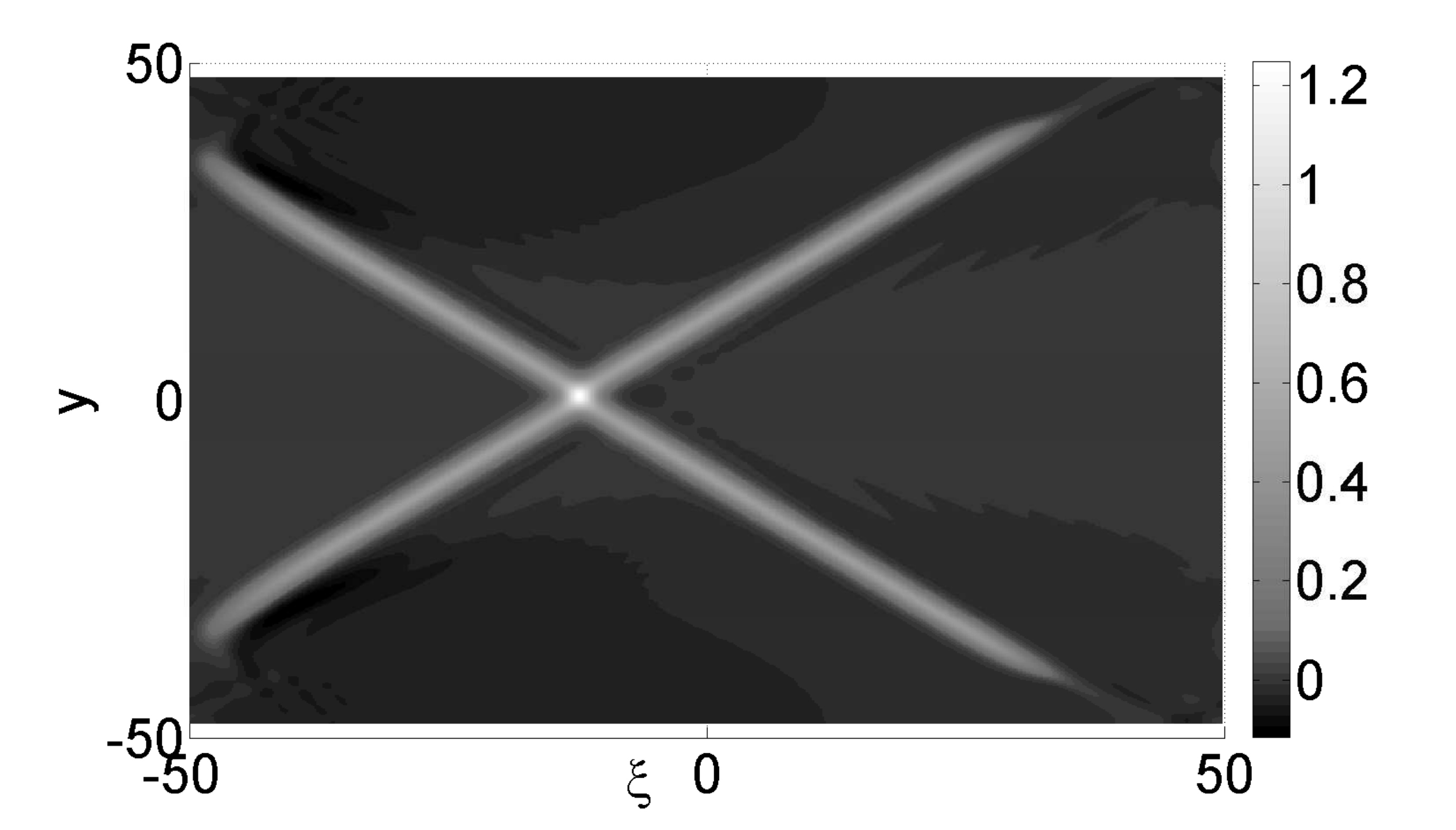}
\caption{Short Stem, $k_{3}=10^{-1}$, $\epsilon=.1$}
\end{subfigure}
\caption{Short Stem X-Waves: $\tau=1/\epsilon$}
\label{fig:xwaveshort}
\end{figure}

\begin{figure}[!h]
\centering
\begin{subfigure}[b]{.49\textwidth}
\includegraphics[width=1\textwidth]{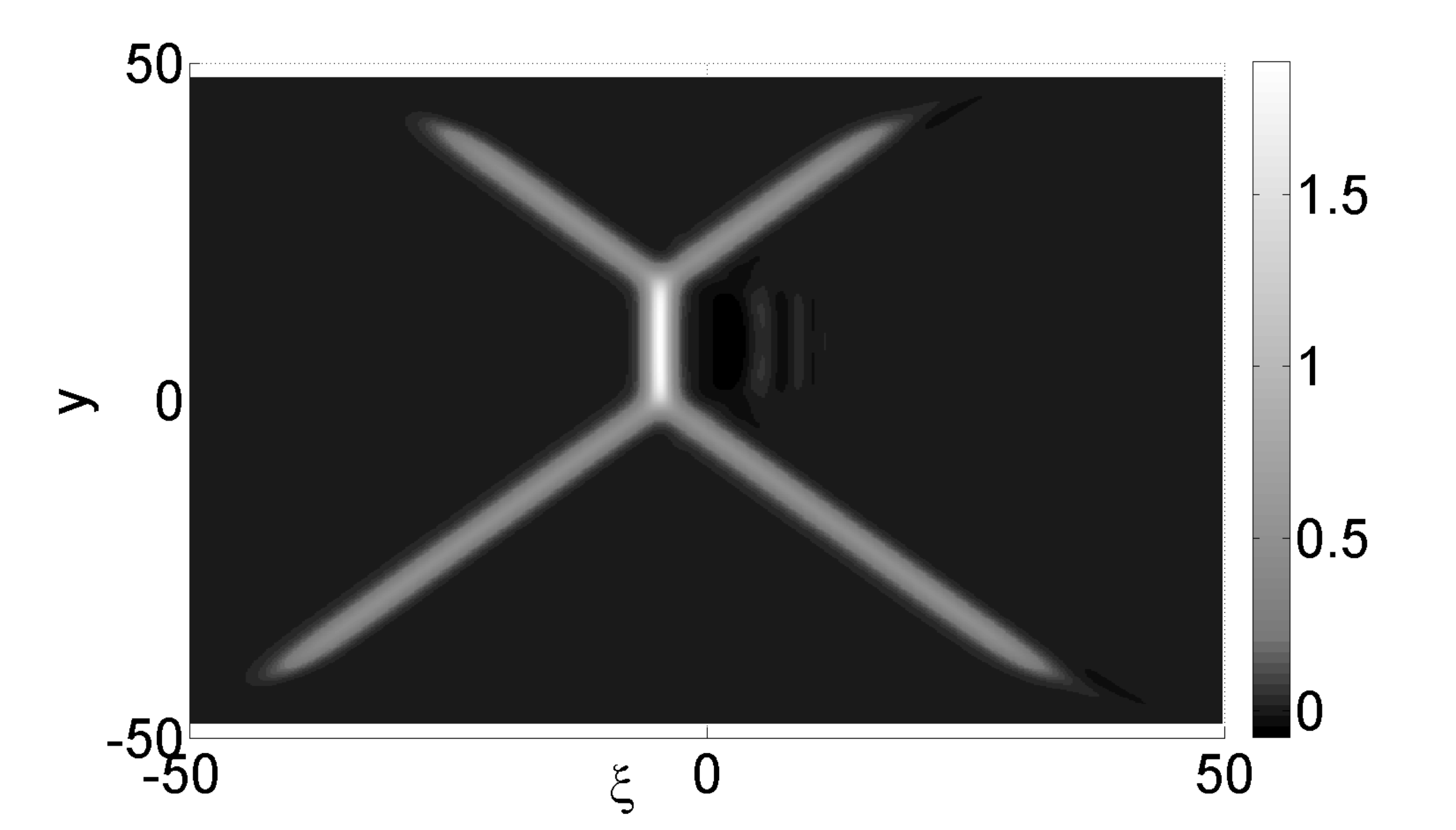}
\caption{Long Stem, $k_{3}=10^{-8}$, $\epsilon = .2$}
\end{subfigure}
\begin{subfigure}[b]{.49\textwidth}
\includegraphics[width=1\textwidth]{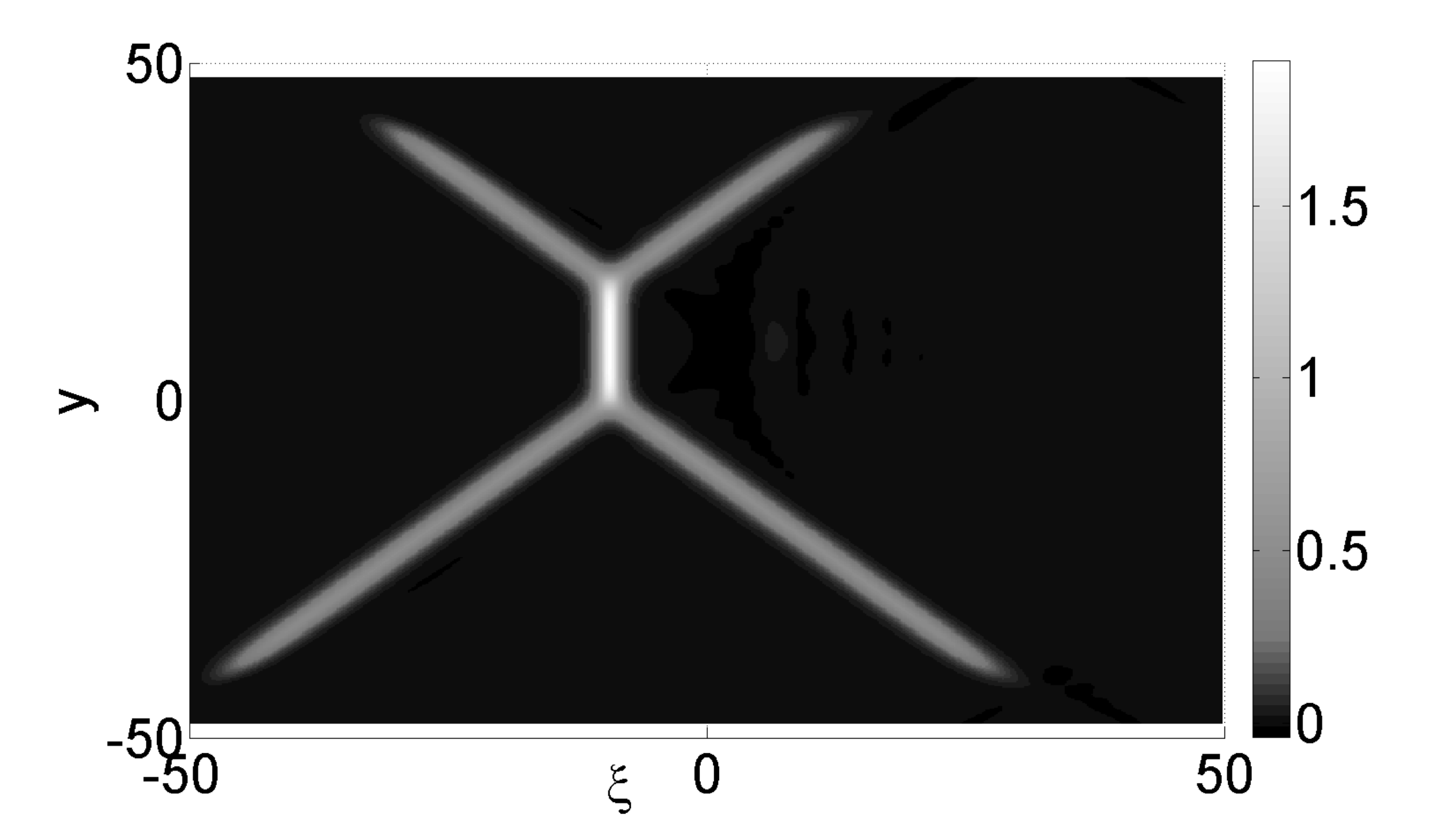}
\caption{Long Stem, $k_{3}=10^{-8}$, $\epsilon = .1$}
\end{subfigure}
\caption{Long Stem X-Waves: $\tau=1/\epsilon$}
\label{fig:xwavelong}
\end{figure}

\begin{figure}[!h]
\centering
\begin{subfigure}[b]{.49\textwidth}
\includegraphics[width=1\textwidth]{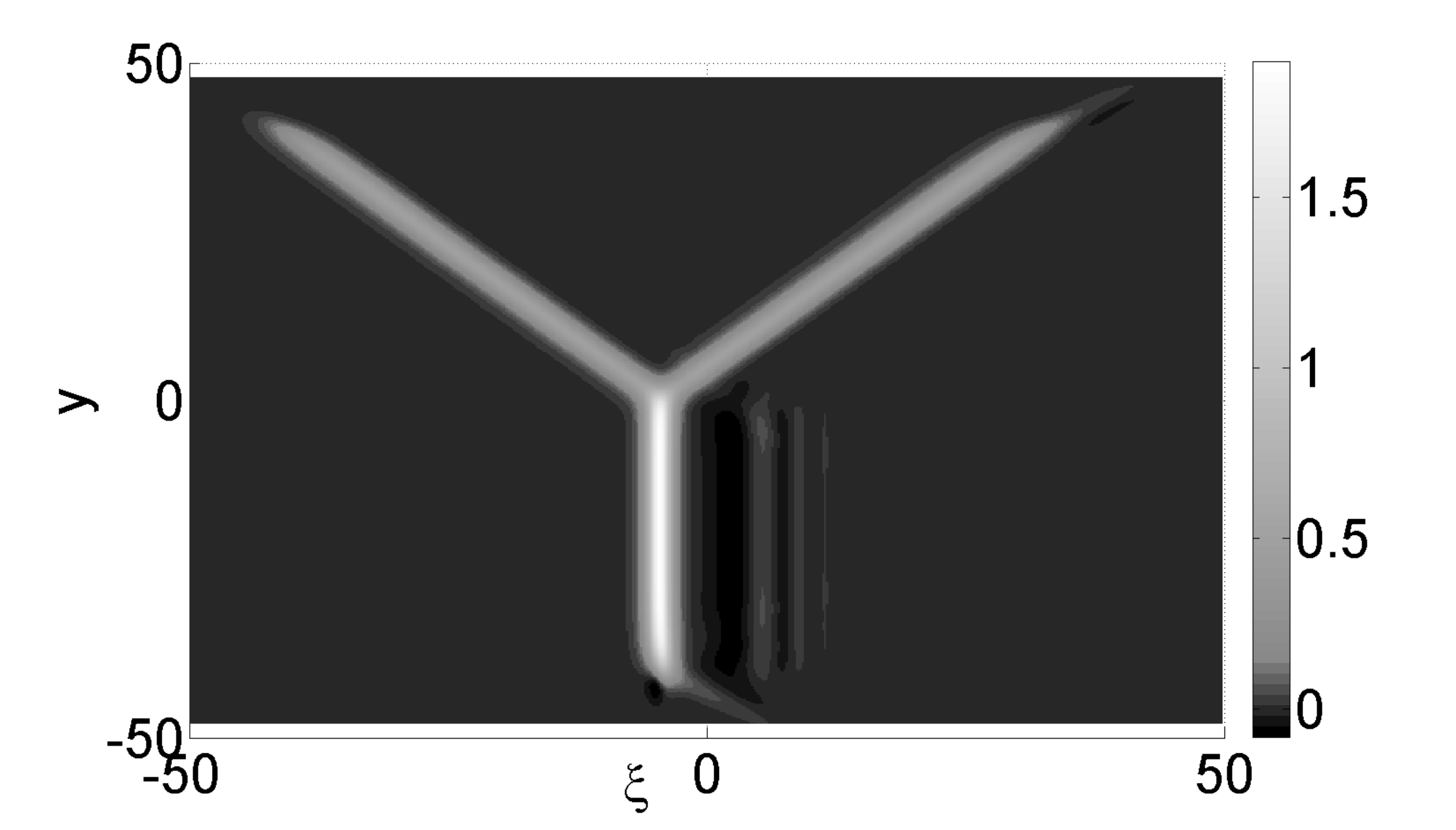}
\caption{Y Junction: $k=1, \epsilon = .2$}
\label{fig:yjuncep2}
\end{subfigure}
\begin{subfigure}[b]{.49\textwidth}
\includegraphics[width=1\textwidth]{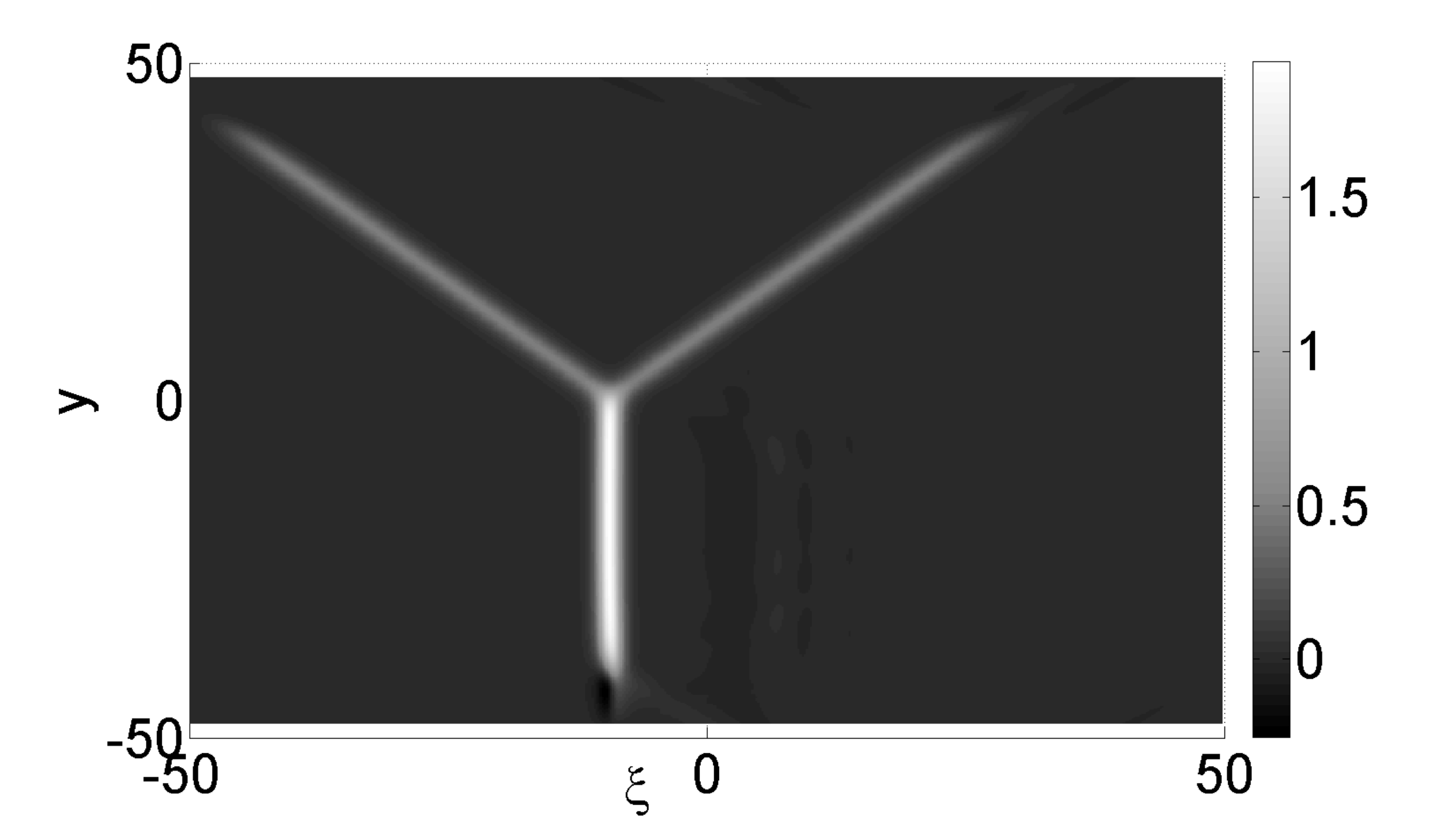}
\caption{Y Junction: $k=1, \epsilon = .1$}
\label{fig:yjuncep1}
\end{subfigure}
\caption{Y-Junctions: $k=1, ~\tau=1/\epsilon$}
\label{fig:yjuncs}
\end{figure}

\noindent In each figure, the initial conditions were chosen as $w(\xi,y,0)$ and $w_{\tau}(\xi,y,0)$ as defined earlier, see \eqref{omega} and \eqref{f_in_Wr}.  The X-wave solutions in Figures \ref{fig:xwaveshort} and  \ref{fig:xwavelong} correspond to the classical two-soliton solutions discussed in \cite{aseg} and \cite{milestwo}.  In the language used earlier, we choose the matrix of coefficients $B$ with entries $b_{ij}$ as, see \eqref{omega} and \eqref{f_in_Wr}, 
\[
B = \left(\ba{rrrr} 1 & b_{12} & 0 & 0 \\ 0 & 0 & 1 & b_{24} \ea\right).
\]
This gives the function $\Omega(\xi,y,\tau)$ as 
\begin{align*}
\Omega = & (k_{3}-k_{1})e^{\theta_{1}+\theta_{3}} + (k_{4}-k_{1})b_{24}e^{\theta_{1}+\theta_{4}} \\
& + (k_{3}-k_{2})b_{12}e^{\theta_{2}+\theta_{3}} + (k_{4}-k_{2})b_{12}b_{24}e^{\theta_{2}+\theta_{4}}.
\end{align*}
We then choose $b_{12} = \frac{k_{4}-k_{1}}{k_{4}-k_{2}}, ~ b_{24} = \frac{k_{3}-k_{2}}{k_{4}-k_{2}}$, and we set $k_{1}=-k_{4}$, $k_{2}=-k_{3}$, $k_{3}>0$, and $k_{4}=1+k_{3}$.  Given this choice of coefficients $b_{ij}$ and $k_{l}$, by letting $k_{3}\rightarrow 0$, one can let the length of the stem grow as shown in Figure \ref{fig:xwavelong}.
Note that the radiation is more significant in Figures \ref{fig:xwaveshort}a,  \ref{fig:xwavelong}a  where $\epsilon=0.2$ than  Figures \ref{fig:xwaveshort}b,  \ref{fig:xwavelong}b where $\epsilon=0.1$.

A different case can be investigated by setting $k_{3}=k_{2}$. Then $\Omega$ becomes
\[
\Omega = (k_{2}-k_{1})\left(e^{\theta_{1}+\theta_{2}} + \frac{k_{4}-k_{1}}{k_{2}-k_{1}}b_{24}e^{\theta_{1}+\theta_{4}}+\frac{k_{4}-k_{2}}{k_{2}-k_{1}}b_{12}b_{24}e^{\theta_{2}+\theta_{4}}\right).
\]
If we then choose
\[
b_{24} = \frac{k_{2}-k_{1}}{k_{4}-k_{1}}, ~ b_{12} = \frac{k_{4}-k_{1}}{k_{4}-k_{2}}, 
\] 
and relabel $\theta_{4}$ as $\theta_{3}$, we then get $\Omega = (k_{2}-k_{1})\tilde{\Omega}$, where
\[
\tilde{\Omega} = e^{\theta_{1}+\theta_{2}} +  e^{\theta_{2}+\theta_{3}} +  e^{\theta_{1}+\theta_{3}}.
\]
We note that 
\[
\p_{\xi}^{2}\ln((k_{2}-k_{1})\tilde{\Omega}) = \p_{\xi}^{2}\ln(\tilde{\Omega}),
\]
and so from this point on we use $w = 2 \p_{\xi}^{2}\ln(\tilde{\Omega})$ as a solution to the KP equation.  The function $\tilde{\Omega}$ corresponds to the $B$ matrix 
\[
B = \left(\ba{rrr} 1 & 0 & -1  \\ 0 & 1 & 1 \ea\right),
\]
which shows that the $X$-wave has passed to a $M=3$, $N=2$ web-solution.  This is an example of the Y-junction.  We choose the parameters $k_{j}$ such that 
\begin{equation}
k_{1}=-\frac{1+k}{2}, ~ k_{2}=\frac{1-k}{2}, ~ \mbox{and} ~ k_{3}=\frac{1+k}{2}, 
\label{yjuncparams}
\end{equation}
where $0\leq k \leq 1$.  
Again, we note that the radiation is more significant in Figure \ref{fig:yjuncs}a where $\epsilon=0.2$ than  Figure \ref{fig:yjuncs}b where $\epsilon=0.1$.

These Y-junctions have a ray along $y \rightarrow -\infty$ with maximum amplitude $|u_{max}|=\frac{(1+k)^{2}}{2}$, another ray along the angle $\mbox{arctan}(-k)$ of amplitude $1/2$, and finally a ray of amplitude $k^{2}/2$ moving at an angle of $\pi/4$ where the angles are measured relative to the y-axis.  Thus as $k \rightarrow 1$, the KP equation predicts a four-fold amplification of the ray along $y\rightarrow -\infty$ relative to the amplitude of the ray along the angle $\mbox{arctan}(-k)$.  Likewise, as we vary $k$, we move between a one-dimensional line soliton solution when k=0  up to a Y-junction with maximal amplification ratio of a factor four when $k=1$, see Figure \ref{fig:yjuncs}.  Alternatively we can use the formalism in \cite{satsuma} to get these initial conditions.

The figures represent numerical simulations run on time scales long enough to allow the next order terms that distinguish the BL equation from the KP equation to have a significant impact.  This effect manifests itself in weak dispersive tails indicated in the figures.  However, in all six figures, it can be seen that the primary wave remains essentially unchanged over time scales that are the reciprocal of the order of magnitude of the perturbation.  We also point out that the figures show that as $\epsilon$ decreases, or as the water becomes shallower, the amount of radiation decreases.  These figures indicate that the web-solutions studied here are stable with respect to evolution in the BL equation.  Hence this result provides quantitative evidence that the class of web-solutions examined in this paper are robust with respect to the perturbations introduced by the BL equation.  
\subsection{Amplification Ratio for the Y-Junction in the Benney--Luke Equation}
The problem of the amplification ratio of the stem in a Y-junction is an interesting issue that was first studied in \cite{milestwo} in the context of the KP equation.  We now discuss how the BL equation affects the amplification ratio, and how this class of KP solutions behaves under the influence of the BL equation.  

\begin{figure}[!h]
\begin{center}
\includegraphics[width=.6\textwidth]{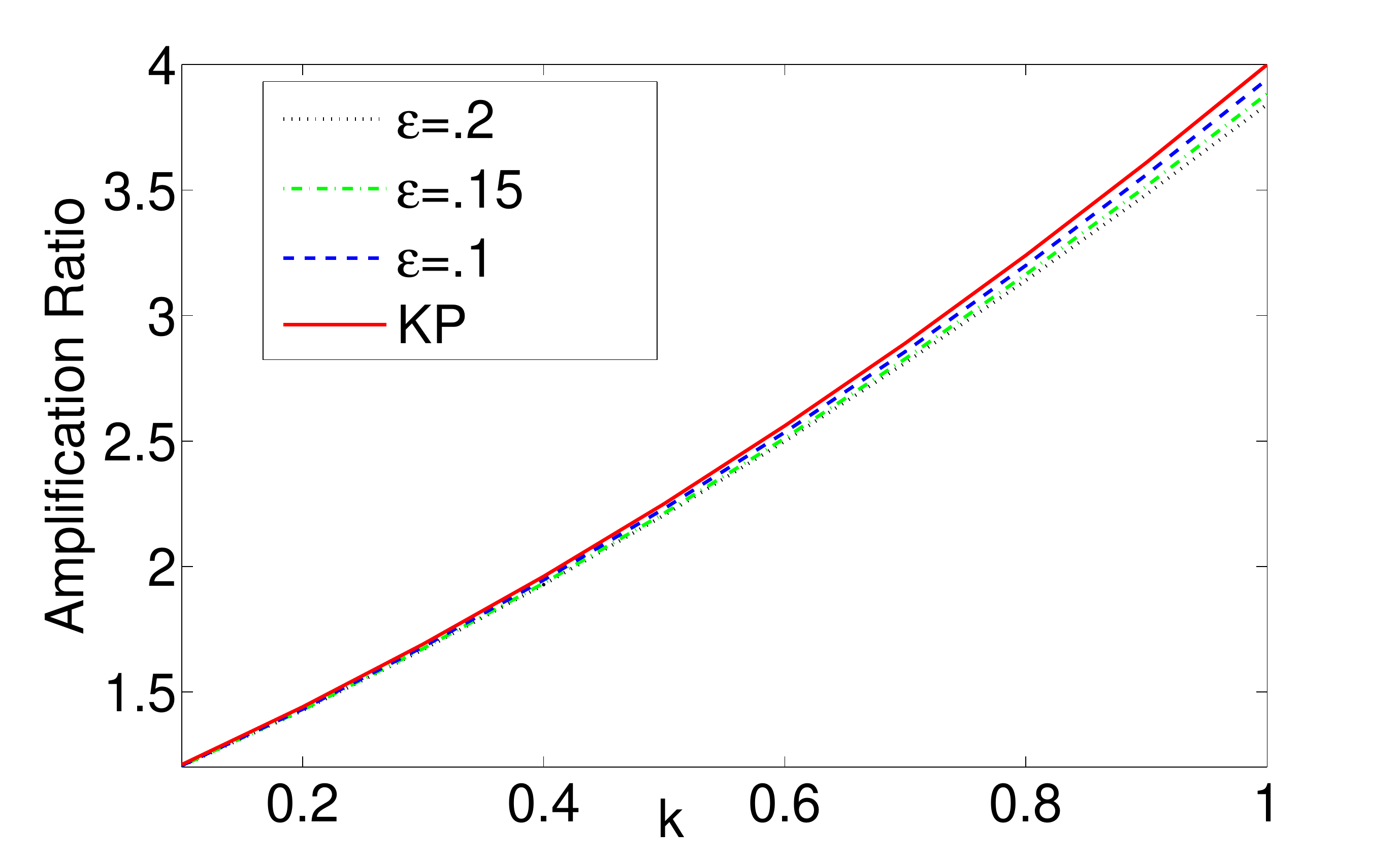}
\caption{Amplification Ratio of the ``Mach Stem" in the BL equation.}
\label{fig:asymp}
\end{center}
\end{figure}

As indicated above, taking a Y-junction with parameter values given by \eqref{yjuncparams}, and its corresponding derivative with respect to time, as Cauchy data, Figure \ref{fig:asymp} shows the amplification ratio (vertical axis) of a Y-junction in the BL equation for  different values of $\epsilon$ and $k$.  For each curve, the simulations were run for times up to $\tau = \frac{1}{\epsilon}$, thus allowing time for the BL equation to have an asymptotically significant impact.  The domain size was chosen to be $L_{\xi}=50$ and $L_{y}=48$ with 512 modes used for the pseudo-spectral approximation.  The size of the rays of the Y-junction were measured at $L_{y}=\pm 30$ so as to avoid any effects from the windowing.  In order to find the wave amplitude, we averaged the height of the wave over seven mesh points around the grid point corresponding to $L_{y} = \pm 30$; this minimizes the impact of any spurious oscillations due to the numerical method.  Finally, we chose $k=.1, \cdots, 1$ in intervals of a tenth.  Figures \ref{fig:yjuncep2} and \ref{fig:yjuncep1} are plots of the case $k=1$ for $\epsilon = .2$ and $\epsilon = .1$ respectively.      

We also note that the phase, or speed correction, along the ray $y\rightarrow -\infty$ is given by $\p_{T}\theta^{(0)}_{13} = \frac{(1+k)^{4}}{96}$ so that the speed, see \eqref{speedformula}, of the ray in the far field should be $-\left(1+ \epsilon \frac{(1+k)^{4}}{96}\right)$, where the negative sign indicates the ray moves to the left.  While this correction goes into the windowing approximation, as a consistency check, we also numerically computed the speed of the ray at $y=-30$ for $\epsilon = .1$ and $k=1$.  After averaging to take into account errors from discretization, we got a computed speed of $-1.0090$, while the predicted speed from the asymptotics is $-1.0167$, so the agreement is very good.    

Figure \ref{fig:asymp} indicates that the numerically generated solutions tend to the exact results for the KP equation as $\epsilon$ decreases and shows that increasing $\epsilon$ decreases the amplification ratio for each $k$.  As indicated above, we note that larger values of $\epsilon$ introduce more dispersive radiation.  This also implies lower pulse heights due to energy conservation which in turn suggests that there will be a somewhat smaller amplification ratio than four at $k=1$ of $O(\epsilon)$ ({\it i.e.} about 3.9 for $\epsilon=.1$), but still much larger than linear theory. This result also agrees with the fact that the BL equation contains the KP equation as an asymptotic limit.  

The variation of the amplification as a function of $\epsilon$ holds uniformly for all $k$.  This indicates that the numerics is well-behaved as we vary the shape of the initial conditions.   Therefore, we find that the combination of asymptotic evaluation of the far-field, pseudo-spectral method, ETDRK4, and windowing is an effective way to approach problems with non-decaying solutions.   
\subsection{Dissipation in the Y-Junction}

\begin{figure}[!h]
\centering
\begin{subfigure}[b]{.49\textwidth}
\includegraphics[width=.75\textwidth]{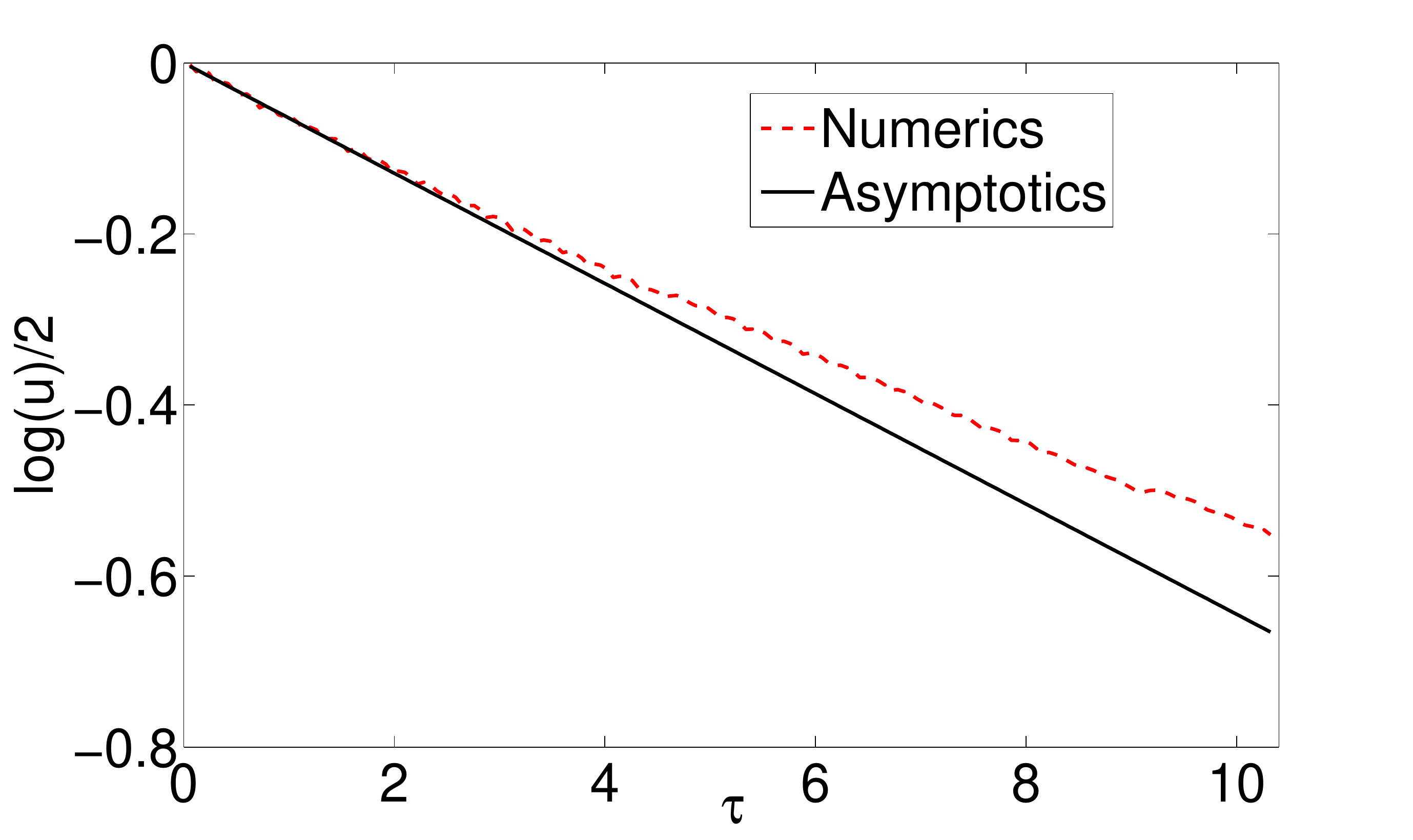}
\caption{Amplitude Decay Comparison}
\end{subfigure}
\begin{subfigure}[b]{.49\textwidth}
\includegraphics[width=1\textwidth]{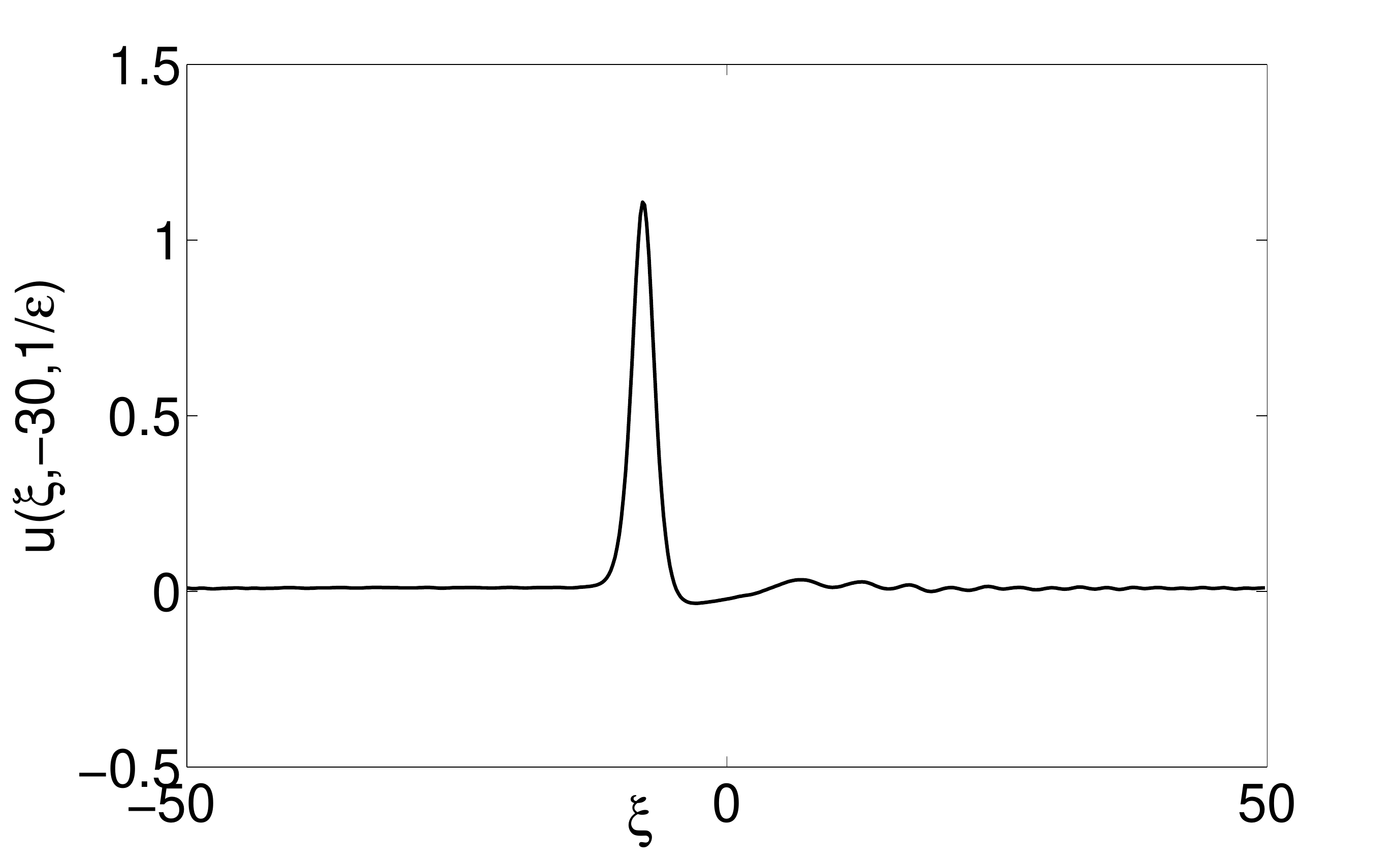}
\caption{Solution profile at $y=-30$}
\end{subfigure}
\caption{$\gamma=2$, $\epsilon=.1$}
\label{fig:ampdecay}
\end{figure}

We next provide numerical results corresponding to the linear dissipation model discussed earlier.  As a leading order solution, we use the Y-junction from the previous section with $k=1$ so that $k_{13}=2$, and we set the coefficient of dissipation $\gamma = 2$.  However, we also take into account the dissipation in the far-field using the asymptotic method shown earlier; this is how we find $u_{asym}$ and $v_{asym}$.  The size of the domain is $L_{\xi}=50$ and $L_{y}=48$, with 512 modes used in the pseudo-spectral method.  We measure the maximum height of the numerical solution in the region $-30\leq y \leq 30$ in order to avoid any effects from the windowing.  The plot in Figure \ref{fig:ampdecay} gives a log plot of the amplitude decay.  

Thus, at $\tau = 1/\epsilon$, the asymptotic theory predicts that the maximum amplitude of the ray should be $2\cdot e^{-.66} =1.033$, while the numerics gives the maximum amplitude as $2\cdot e^{-.55} =1.15$.  Thus the numerics confirms the asymptotic prediction since the discrepancy between the two results is $\mathcal{O}(\epsilon)$.  
We also look at a cross-section of the ray at $y=-30$ in Figure \ref{fig:ampdecay}.  As shown in the figure, to the right of the peak a shelf forms.  This is consistent with the dissipative theory of KdV for linear dissipative models of wave amplitude \cite{ak}; this is another novel feature of web-solutions not yet reported in the literature.    

\section*{Acknowledgements}

This research was partially supported by the National Science Foundation under grant DMS-0905779. We thank Douglas Baldwin for many stimulating and enlightening conversations. 

\section*{Appendix}

\subsection*{Basic Properties of the Average}

While this paper does not present rigorous results, we include some analytical details that add support to the formal arguments presented in the paper.  Supposing $f(\xi,y) \in L^{1}(\mathbb{R}\times [-L,L])$, which is to say that $f(\xi,y)$ is in $L^{1}(\mathbb{R})$ with respect to $\xi$ and $L^{1}([-L,L])$ in $y$, then the quantity 
\[
 \int^{L}_{-L}\int_{\mathbb{R}} f(\xi,y) dy d\xi 
\]
is well-defined, and by the Fubini-Tonelli theorem, and assuming the limits exist, we have that 
\[
\ds{\lim_{L\rightarrow \infty} \int_{\mathbb{R}}} \frac{1}{L}\int^{L}_{-L} f(\xi,y) dy d\xi =  \ds{\lim_{L \rightarrow \infty}} \frac{1}{L}\int_{-L}^{L}\int_{\mathbb{R}} f(\xi,y) d\xi dy. 
\]
We also have that if $f(\xi,y)$ and $f_{\xi}(\xi,y)$ are in $ L^{1}(\mathbb{R}\times [-L,L])$, then 
\[
 \left<  \int_{\mathbb{R}} f_{\xi}(\xi,\cdot) d\xi \right>_{y} = \lim_{L \rightarrow \infty} \frac{1}{L}\int_{-L}^{L}\int_{\mathbb{R}} f_{\xi}(\xi,y) d\xi dy = 0. 
\]
Likewise, we want to establish conditions for when 
\[
  \p_{\tau} \left< \int_{\mathbb{R}} f(\xi,\cdot,\tau) d\xi \right>_{y} = 
 \left<  \int_{\mathbb{R}} \p_{\tau} f(\xi,\cdot,\tau) d\xi \right>_{y}, 
\]
which is equivalent to showing 
\[
\lim_{h\rightarrow 0} \lim_{L\rightarrow \infty}  \frac{1}{2L}\int_{-L}^{L}\int_{\mathbb{R}} \left|f_{\tau,h}(\xi,y,\tau)-f_{\tau}(x,y,\tau) \right|d\xi dy = 0, 
\]
where 
\[
f_{\tau,h}(\xi,y,\tau)= \frac{f(\xi,y,\tau+h)-f(\xi,y,\tau)}{h}. 
\]
We assume that $f(\xi,y,\tau),~ \p_{\tau}f(\xi,y,\tau) \in L^{1}(\mathbb{R}\times [-L,L])$ on some interval $\tau \in (\tau_{1},\tau_{2})$.  Then we have, using dominated convergence, for any $L>0$, that 
\[
\lim_{h\rightarrow 0} \frac{1}{2L}\int_{-L}^{L}\int_{\mathbb{R}} \left|f_{\tau,h}(\xi,y,\tau)-f_{\tau}(x,y,\tau) \right|d\xi dy = 0. 
\]
Define the function $\tilde{f}(L,h)$ such that 
\[
\tilde{f}(L,h) = \frac{1}{2L}\int_{-L}^{L}\int_{\mathbb{R}} \left|f_{\tau,h}(\xi,y,\tau)-f_{\tau}(x,y,\tau) \right|d\xi dy.
\]
We assume $\sup_{L>L^{\ast}}\tilde{f}(L,h)<\infty$ for some $L^{\ast}>0$.  Then, for all $\epsilon > 0$, there must be some value $\tilde{L}$ such that  
\[
\limsup_{L \rightarrow \infty}  \frac{1}{2L}\int_{-L}^{L}\int_{\mathbb{R}} \left|f_{\tau,h}(\xi,y,\tau)-f_{\tau}(x,y,\tau) \right|d\xi dy \leq  \tilde{f}(\tilde{L},h)+ \epsilon. 
\]
Taking the limit on both sides as $h\rightarrow 0$, and noting that $\epsilon > 0$ is arbitrary shows 
\[
\lim_{h\rightarrow 0} \limsup_{L \rightarrow \infty}  \frac{1}{2L}\int_{-L}^{L}\int_{\mathbb{R}} \left|f_{\tau,h}(\xi,y,\tau)-f_{\tau}(x,y,\tau) \right|d\xi dy =0. 
\]
Since we are only working with positive quantities, this shows that 
\[
\lim_{h\rightarrow 0} \lim_{L \rightarrow \infty}  \frac{1}{2L}\int_{-L}^{L}\int_{\mathbb{R}} \left|f_{\tau,h}(\xi,y,\tau)-f_{\tau}(x,y,\tau) \right|d\xi dy =0, 
\]
and the result is established.  Thus we have found a class of functions for which the analytic operations performed throughout the paper are valid.  We believe the web solutions and their perturbations satisfy the conditions listed above, but it is beyond the scope of the paper to prove as such.   


\subsection*{Conservation of Energy for KP Web Solutions}
In this section, we show the condition (\ref{finalconclusion}) holds for web solutions.  We also establish a number of properties about the web-solutions with regards to the average used in defining the energy.  Again we point out that if (\ref{finalconclusion}) holds, then from 
\[
 \left<  \ds{\int_{\mathbb{R}}} ww_{\tau} d\xi \right>_{y}  = \left< \ds{\int_{\mathbb{R}}} wK(w) d\xi \right>_{y} = \frac{3}{4}  \left<  \ds{\int_{\mathbb{R}}} w \p^{-1}_{\xi} w_{yy} d\xi  \right>_{y},
\]
where we have assumed perfect derivatives in $\xi$ cancel, $\p_{\tau}\mathcal{E}(w)=0$.  To get the derivative of the energy to vanish then, we must prove (\ref{finalconclusion}) for $u=w$.  To do this, using integration by parts, we find 
\begin{align}
\frac{1}{L}\int^{L}_{-L}\int_{\mathbb{R}}w\p^{-1}_{\xi}w_{yy} dy d\xi = & \int_{\mathbb{R}}\frac{1}{L}\left.w\p^{-1}_{\xi}w_{y}\right|^{y=L}_{-L} d\xi \label{enervar} \\ 
& - \frac{1}{L}\int^{L}_{-L}\int_{\mathbb{R}}w_{y}\p^{-1}_{\xi}w_{y} dy d\xi. \nonumber
\end{align}
Since we have that $w_{y}\p^{-1}_{\xi}w_{y}=\frac{1}{2}\p_{\xi} (\p^{-1}_{\xi}w_{y})^2$ and we expect that $(\p^{-1}_{\xi}w_{y})^2\rightarrow 0$, \cf  \cite{av}, and $\frac{1}{L}\left.w\p^{-1}_{\xi}w_{y}\right|^{y=L}_{-L} \rightarrow 0$ as $|L| \rightarrow \infty$, then the result should follow. 

To prove (\ref{finalconclusion}) rigorously, we begin by showing that $w$ is always positive, since this allows us to introduce notation and simplify some computations.  To show $w\geq 0$, note that $w$ can be written as 
\[
w = \frac{\Omega_{\xi \xi }\Omega - \Omega^{2}_{\xi}}{\Omega^{2}}, 
\]
and that we can write $\Omega$ in the form $\Omega = \mbox{det}(BEK)$, where $B$ is a $N \times M$ matrix with entries $b_{ij}$, $E$ is an $M\times M$ diagonal matrix with $E_{jj} = e^{\theta_{j}}$, and $K$ is an $M \times N$ matrix with entries $K_{ij}=k_{i}^{(j-1)}$ (\cf \cite{chak}).  Using the Cauchy-Binet Theorem, $\Omega$ can be written as 
\[
\Omega = \sum_{S \in \left(\left<M\right> \over N \right)} \mbox{det}(B_{<N>,S})\mbox{det}((EK)_{S,<N>}), 
\]
where $\left<M\right> = \left\{1, 2, \cdots, M \right\}$ and $\left( \frac{\left<M\right>}{N} \right)$ represents all possible $N$ permutations of the $M$ numbers in $\left<M\right>$.  The symbol $B_{<N>,S}$ denotes the $N\times N$ matrix formed from the $S=\left\{S(1), S(2), \cdots, S(N) \right\}$ columns of $B$.  Likewise, $(EK)_{S,<N>}$ is the $N \times N$ matrix formed from the $S$ rows of the matrix $EK$.  By construction it is assumed that $\mbox{det}(B_{<N>,S}) \geq 0$, and that not all of these terms can be identically zero.  Likewise, for a given $S$, it follows that 
\[
\mbox{det}((EK)_{S,<N>}) = \exp\left(\sum_{l=1}^{N}\theta_{S(l)}\right)\prod_{1\leq i < j \leq N}(k_{S(j)}-k_{S(i)}). 
\]
Therefore $\Omega > 0$ since $k_{S(j)}>k_{S(i)}$ when $j>i$.  Letting 
\[
\Delta_{S} = \mbox{det}(B_{<N>,S})\prod_{1\leq i < j \leq N}(k_{S(j)}-k_{S(i)}), 
\]
it is straightforward to show that $\Omega_{\xi \xi}\Omega - (\Omega_{\xi})^{2}$ is equal to
\[
\sum_{S,S'} \Delta_{S}\Delta_{S'}\exp\left(\sum_{j=1}^{N}\theta_{S(j)} + \theta_{S'(j)} \right)\sum_{l=1}^{N}k_{S(l)}\sum_{n=1}^{N}\left(k_{S(n)}-k_{S'(n)}\right). 
\]
In the sums, if we have the pair $(S,S')$, then we must also have the pair $(S',S)$, and combining these terms gives 
\[
\Delta_{S}\Delta_{S'}\exp\left(\sum_{j=1}^{N}\theta_{S(j)} + \theta_{S'(j)} \right)\left(\sum_{l=1}^{N}\left(k_{S(l)}-k_{S'(l)}\right)\right)^{2},
\]
and therefore $w \geq 0$.  

Since $\Omega$ is smooth in all its arguments, $w$ is as well, and we then have
\[
\left.\int_{L_{1}}^{L_{2}} w(\xi,y,\tau,T) d\xi = 2\p_{\xi}\ln \Omega \right|^{L_{2}}_{\xi = L_{1}}. 
\]  
We need to determine the behavior of $\Omega_{\xi}/\Omega$ as $L_{2}\rightarrow \infty$ and $L_{1}\rightarrow -\infty$.  In the case that $L_{2}\rightarrow \infty$, there is an $S$, say $S_{max} = \left\{k_{M - N + 1}, \cdots, k_{M} \right\}$, such that for $\xi \gg 1$,
\[
\exp\left(\sum_{j=1}^{N}\theta_{S_{max}(j)} \right) \gg \exp\left(\sum_{j=1}^{N}\theta_{S(j)}\right). 
\]
We write $\Omega_{\xi}/\Omega$ as 
\[
\frac{\Omega_{\xi}}{\Omega} = \sum_{l=1}^{N}k_{S_{max}(l)} + o(1),~\xi \rightarrow \infty, 
\]  
where we have divided through by $\Delta_{S_{max}}\exp\left(\sum_{j=1}^{N}\theta_{S_{max}(j)} \right)$ to isolate the leading order term.  An identical argument shows that 
\[
\frac{\Omega_{\xi}}{\Omega} = \sum_{l=1}^{N}k_{S_{min}(l)} + o(1),~\xi \rightarrow -\infty, 
\]
where $S_{min}=\left\{1, 2, \cdots, N \right\}$.  Therefore, we have shown that 
\[
\int_{-\infty}^{\infty} w(\xi,y,\tau,T) d\xi = 2 \left(\sum_{l=1}^{N}k_{S_{max}(l)} - \sum_{l=1}^{N}k_{S_{min}(l)}\right). 
\]  
This result also establishes that the average $\left<\int_{\mathbb{R}} w d\xi\right>_{y}$ is well defined.  Using the inequality
\[
\frac{1}{2L}\int_{-L}^{L}\int_{\mathbb{R}}w^{2}(\xi,y,\tau,T)d\xi dy \leq \sup_{\mathbb{R}^{2}}|w|~ \frac{1}{2L}\int_{-L}^{L}\int_{\mathbb{R}}w(\xi,y,\tau,T)d\xi dy \nonumber
\] 
yields, noting that $\sup_{\mathbb{R}^{2}}|w| < \infty$,
\[
\limsup_{L \rightarrow \infty} \frac{1}{2L}\int_{-L}^{L}\int_{\mathbb{R}}w^{2}(\xi,y,\tau,T)d\xi dy < \infty. 
\]
However, establishing that the limit exists, or that $\mathcal{E}(w)$ is rigorously well defined for any web solution is technically more demanding and beyond the scope of this paper.  Therefore, we consider the energy $\mathcal{E}(w)$ formally.    

Returning to evaluating \eqref{enervar}, for the second term on the right hand side, using $w_{y}\p^{-1}_{\xi}w_{y}=\frac{1}{2}\p_{\xi} (\p^{-1}_{\xi}w_{y})^2$, we see we need to evaluate
\[
\lim_{L_{1},L_{2} \rightarrow \infty} \left.\left(\p^{-1}_{\xi}w_{y}\right)^{2}\right|^{L_{2}}_{\xi = -L_{1}}. 
\]
We have that 
\[
\p^{-1}_{\xi}w_{y} = 2\p_{y\xi}\ln \Omega - \lim_{\xi \rightarrow -\infty}\p_{y\xi}\ln \Omega - \lim_{\xi \rightarrow \infty}\p_{y\xi}\ln \Omega. 
\]
Given $\p_{y \xi} \ln \Omega = \ds{\frac{\Omega_{\xi y}\Omega-\Omega_{\xi}\Omega_{y}}{\Omega^{2}}}$, we write $\Omega_{\xi y}\Omega-\Omega_{\xi}\Omega_{y}$ as 
\[
\sum_{S,S'} \Delta_{S}\Delta_{S'}\exp\left(\sum_{j}\theta_{S(j)}+\theta_{S'(j)}\right) \sum_{j=1}^{N}k_{S(j)}\left(\sum_{j=1}^{N}k^{2}_{S(j)} - k^{2}_{S'(j)}\right).
\] 
Thus, in the numerator of $\ds{\frac{\Omega_{\xi y}\Omega-\Omega_{\xi}\Omega_{y}}{\Omega^{2}}}$, terms in the sum where $S=S'$ cancel, which does not happen in the denominator.  Thus, as $\xi \rightarrow \infty$, the dominant behavior of the denominator, $\Omega^{2}$, is $e^{2\theta_{S_{max}}}$, which is exponentially larger than any term in the numerator.  A similar argument holds for $\xi \rightarrow -\infty$, and thus $\lim_{|\xi| \rightarrow \infty} \p_{y \xi}\ln \Omega = 0$, and therefore, for any $L>0$, 
\[
\int^{L}_{-L}\int_{\mathbb{R}}w_{y}\p^{-1}_{\xi}w_{y} dy d\xi = 0.
\]

Finally, we now need to show that 
\[
\lim_{L\rightarrow \infty} \int_{\mathbb{R}}\frac{1}{L}\left.w\p^{-1}_{\xi}w_{y}\right|^{y=L}_{-L} d\xi = 0. 
\]
It is straightforward to show that $\left|\p^{-1}_{\xi}w_{y}\right|$ is uniformly bounded above throughout the $\xi - y$ plane, say $\left|\p^{-1}_{\xi}w_{y}\right|\leq C$, so that  $\left|w\p^{-1}_{\xi}w_{y}\right| \leq C w$.  Thus
\[
\left| \int_{\mathbb{R}}\frac{1}{L}\left.w\p^{-1}_{\xi}w_{y}\right|^{y=L}_{-L} d\xi\right| \leq \left.\frac{C}{L}( \p_{\xi}\ln \Omega(\xi,L,\tau,T) + \p_{\xi}\ln \Omega(\xi,-L,\tau,T))\right|^{\infty}_{-\infty}, 
\]
which immediately gives the estimate
\[
\left| \int_{\mathbb{R}}\frac{1}{L}\left.w\p^{-1}_{\xi}w_{y}\right|^{y=L}_{-L} d\xi\right| \leq \frac{2C}{L}\left(\sum_{l=1}^{N}k_{S_{max}(l)} - \sum_{l=1}^{N}k_{S_{min}(l)}\right),
\]  
and therefore we obtain the desired result (\ref{finalconclusion}).

\subsection*{Calculation of the pseudo-spectral representation of $\p^{-1}_{\xi}$}

We take the number of interpolation points in the pseudo-spectral method to be $N_{T}$, which is assumed to be an even number, {\it i.e.} $N_{T}=2\tilde{N}$.  Then each interpolation point is given by
\[
\xi_{m} = -L_{\xi} + \frac{2L_{\xi}}{N_{T}}m, ~ y_{n} = -L_{y} + \frac{2L_{y}}{N_{T}}n,  
\] 
with $m=0,\cdots,N_{T}-1$ and $n=0,\cdots,N_{T}-1$.  We define the coefficients $\tilde{a}_{jl}(\tau) = (-1)^{l+j}a_{jl}(\tau)$, so that at the point $(\xi_{m},y_{n})$, using \eqref{dinverseI}, we have 
\begin{eqnarray}
\p^{-1}_{\xi} \tilde{u}(\xi_{m},y_{n}) & = & \frac{L_{\xi}}{i\pi} \sum_{j=-\tilde{N}+1}^{\tilde{N}}~\sum_{l=-\tilde{N}+1, l\neq 0}^{\tilde{N}} \frac{\tilde{a}_{jl}(\tau)}{l}  e^{2 \pi i (lm + jn)/N_{T}} \nonumber \\
&& -~ \frac{L_{\xi}}{i\pi} \sum_{j=-\tilde{N}+1}^{\tilde{N}}~\sum_{l=-\tilde{N}+1, l\neq 0}^{\tilde{N}}\frac{\tilde{a}_{jl}(\tau)}{l}  e^{ 2\pi i jn/N_{T}} \nonumber \\
&&+ ~\xi_{m} \sum_{j=-\tilde{N}+1}^{\tilde{N}} \tilde{a}_{j0}(\tau) e^{ 2\pi i j n/N_{T} }.\nonumber
\end{eqnarray}
We then, for integers $\tilde{j}$ and $\tilde{l}$, compute the discrete inverse Fourier transform
\[
(\p^{-1}_{\xi} \tilde{u})^{\hat{}}_{\tilde{j}, \tilde{l}}  = \frac{1}{N^{2}_{T}}\sum_{m=0}^{N_{T}-1}\sum_{n=0}^{N_{T}-1}\p^{-1}_{\xi} \tilde{u}(\xi_{m},y_{n}) e^{-2\pi i (m \tilde{l} + n \tilde{j})/N_{T}},  
\]
which gives
\begin{eqnarray}
(\p^{-1}_{\xi} \tilde{u})^{\hat{}}_{\tilde{j}, \tilde{l}} & = & \frac{L_{\xi}}{i\pi \tilde{l}}~ \tilde{a}_{\tilde{j}\tilde{l}}(\tau)(1-\delta_{\tilde{l} 0}) -\delta_{\tilde{l}0}\frac{L_{\xi}}{i\pi N_{T}} \sum_{l=-\tilde{N}+1,l\neq 0}^{\tilde{N}} \frac{\tilde{a}_{\tilde{j}l}(\tau)}{l} \nonumber \\
 & & +\frac{\tilde{a}_{\tilde{j} 0}(\tau) }{N_{T}} \sum_{m=0}^{N_{T}-1}\xi_{m} e^{-2\pi i m \tilde{l}/N_{T}}. \nonumber
\end{eqnarray}
If $\tilde{l}=0$, the last sum in this expression becomes $-L_{\xi}$.  Otherwise, if $\tilde{l}\neq 0$, we have 
\begin{eqnarray}
\sum_{m=0}^{N_{T}-1} \xi_{m}  e^{-2\pi i m \tilde{l}/N_{T}} & = & -\frac{L_{\xi}}{i\pi} \p_{\tilde{l}} \sum_{m=0}^{N_{T}-1} e^{-2\pi i m \tilde{l}/N_{T}}, \nonumber \\
& = & -\frac{L_{\xi}}{i\pi} \p_{\tilde{l}}\left( \frac{e^{-2\pi i \tilde{l}} -1}{e^{-2\pi i \tilde{l}/N_{T}}-1}  \right) \nonumber
\end{eqnarray}
where we treat $\tilde{l}$ as a continuous parameter.  After differentiating, we get the final expression
\begin{eqnarray}
(\p^{-1}_{\xi} \tilde{u})^{\hat{}}_{\tilde{j}, \tilde{l}} & = & (1-\delta_{\tilde{l} 0})\frac{L_{\xi}}{i\pi \tilde{l}}~ \tilde{a}_{\tilde{j}\tilde{l}}(\tau) - \delta_{\tilde{l}0}\frac{L_{\xi}}{i\pi N_{T}} \sum_{l=-\tilde{N}+1,l\neq 0}^{\tilde{N}} \frac{\tilde{a}_{\tilde{j}l}(\tau)}{l} \nonumber \\
 & & +  ~(1-\delta_{\tilde{l} 0 })\frac{2L_{\xi} }{N_{T}}\frac{\tilde{a}_{\tilde{j} 0}}{e^{-2\pi i  \tilde{l}/N_{T}} -1} - \delta_{\tilde{l} 0}\frac{L_{\xi}}{N_{T}}, \nonumber
\end{eqnarray}
which is \eqref{dinverseII} after relabeling indices.  

\bibliography{BL4.6.12}
\bibliographystyle{plain}

\end{document}